\documentclass[aps,prd,twocolumn,superscriptaddress,preprintnumbers,nofootinbib,letterpaper,longbibliography]{revtex4-1}

\usepackage{graphicx}
\usepackage{amsmath, amssymb}
\usepackage[export]{adjustbox}
\usepackage[caption=false]{subfig}
\usepackage{svg}
\usepackage{hyperref}
\hypersetup{
  colorlinks  = true,
  urlcolor    = blue,
  linkcolor   = blue,
  citecolor   = red
}
\usepackage[capitalize]{cleveref}
\newcommand{\orcid}[1]{\begingroup
  \hypersetup{hidelinks}\href{https://orcid.org/#1}{\includegraphics[width=10pt]{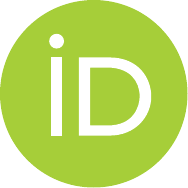}} \endgroup}

\begin{document}

\title{\texorpdfstring{Detector Requirements for Model-Independent Measurements of \\ Ultrahigh Energy Neutrino Cross Sections}{Detector Requirements for Model-Independent Measurements of Ultrahigh Energy Neutrino Cross Sections}}

\author{Ivan Esteban \orcid{0000-0001-5265-2404}\,}
\email{esteban.6@osu.edu}
\affiliation{Center for Cosmology and AstroParticle Physics
  (CCAPP), Ohio State University, Columbus, Ohio 43210}
\affiliation{Department of Physics, Ohio State University, Columbus, Ohio 43210}

\author{Steven Prohira \orcid{0000-0002-8814-6607}\,}    
\email{prohira.1@osu.edu}
\affiliation{Center for Cosmology and AstroParticle Physics
  (CCAPP), Ohio State University, Columbus, Ohio 43210}
\affiliation{Department of Physics, Ohio State University, Columbus, Ohio 43210}

\author{John F. Beacom \orcid{0000-0002-0005-2631}\,}    
\email{beacom.7@osu.edu}
\affiliation{Center for Cosmology and AstroParticle Physics
  (CCAPP), Ohio State University, Columbus, Ohio 43210}
\affiliation{Department of Physics, Ohio State University, Columbus, Ohio 43210}
\affiliation{Department of Astronomy, Ohio State University, Columbus, Ohio 43210}

\date{\today}


\begin{abstract}

The ultrahigh energy range of neutrino physics (above $\sim 10^{7} \, \mathrm{GeV}$), as yet devoid of detections, is an open landscape with challenges to be met and discoveries to be made. Neutrino-nucleon cross sections in that range --- with center-of-momentum energies $\sqrt{s} \gtrsim 4 \, \mathrm{TeV}$ --- are powerful probes of unexplored phenomena. We present a simple and accurate model-independent framework to evaluate how well these cross sections can be measured for an unknown flux and generic detectors.  We also demonstrate how to characterize and compare detector sensitivity.  We show that cross sections can be measured to $\simeq ^{+65}_{-30}$\% precision over $\sqrt{s} \simeq$ 4--140 TeV ($E_\nu = 10^7$--$10^{10}$\, GeV)  with modest energy and angular resolution and $\simeq 10$ events per energy decade.  Many allowed novel-physics models (extra dimensions, leptoquarks, etc.) produce much larger effects.  In the distant future, with $\simeq 100$ events at the highest energies, the precision would be $\simeq 15\%$, probing even QCD saturation effects.

\end{abstract}

\maketitle


\section{Introduction}

\begin{figure}[b]
    \centering
    \includegraphics[width=\columnwidth]{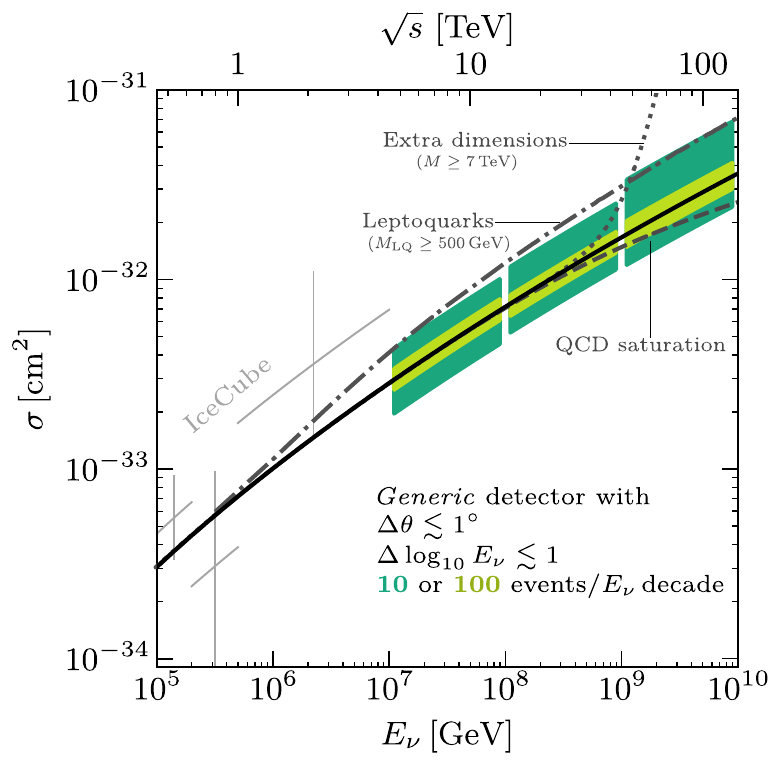}
    \caption{Neutrino-nucleon cross-section sensitivity for UHE astrophysical neutrinos, lower energy data, and novel-physics predictions. See \cref{fig:sensitivity_xsec_dTheta} for different $\Delta \theta$.  The top axis shows the center-of-momentum energy $\sqrt{s}$. \emph{Precise measurements are possible with reasonable detectors and statistics.}}
    \label{fig:sensitivity_xsec}
\end{figure}

New laws of physics are anticipated at high energies. This has stimulated building large colliders with center-of-momentum energies ($\sqrt{s}$) as high as 13.6 TeV. In principle, even higher energies can be probed with ultra-high energy (UHE) particles from astrophysical sources. When these particles interact with nucleons at Earth, they probe large $\sqrt{s} = \sqrt{2 E m_p}$, exceeding $13.6 \, \mathrm{TeV}$ for $E \gtrsim 9 \times 10^{7} \, \mathrm{GeV}$. UHE neutrinos could provide especially powerful tests of novel-physics scenarios, as even subtle new interactions would exceed weak interactions in the cross section with nucleons, $\sigma$. \Cref{fig:sensitivity_xsec} previews our results for three independent energy bins.

Conceptually, measuring $\sigma$ with astrophysical neutrinos is simple, taking advantage of Earth's opacity to high-energy neutrinos to break the degeneracy between the unknown flux and cross section~\cite{Kusenko:2001gj, Anchordoqui:2001cg, Hooper:2002yq, Hussain:2006wg, Borriello:2007cs, Hussain:2007ba, Connolly:2011vc, Marfatia:2015hva}.  Neutrinos reaching the detector through small column densities probe $\phi \sigma$, where $\phi$ is the flux.  Neutrinos reaching the detector through large column densities probe $\phi \sigma e^{-n \sigma L(\theta)}$, where $n$ is the number density of targets and $L(\theta)$ is the traversed distance as a function of zenith angle $\theta$.  At lower energies ($\sqrt{s}$ below a few TeV), IceCube data have been used to measure $\sigma$ through a comparison of downgoing ($\theta < 90^\circ$) and upgoing ($\theta > 90^\circ$) rates~\cite{Bustamante:2017xuy, IceCube:2020rnc}.

\begin{figure}[t]
    \centering
    \includegraphics[width=\columnwidth]{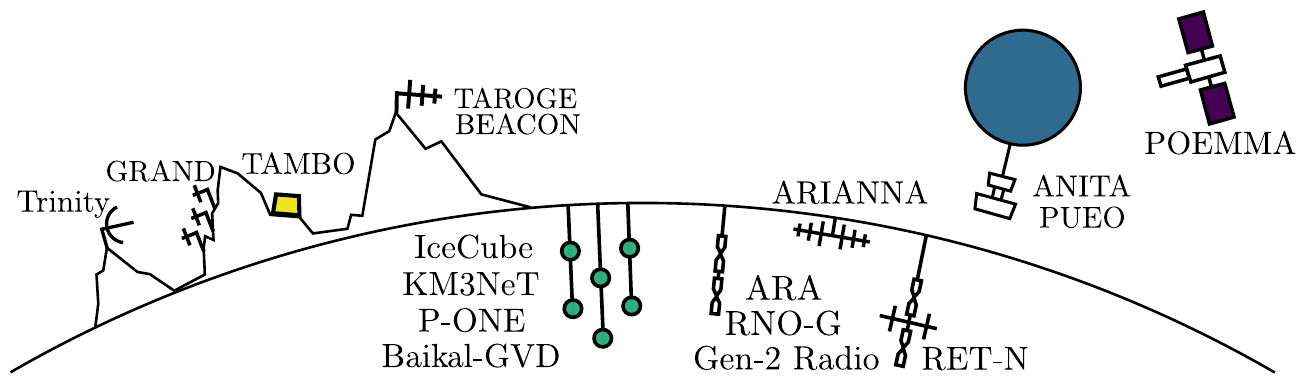}
    \caption{Proposed strategies to detect UHE neutrinos. \emph{The variety guarantees complementary physics opportunities}.}
    \label{fig:experiments}
\end{figure}

Adapting these ideas to UHE neutrinos brings new challenges. Earth attenuation is strong, and most events come from near the horizon. The flux is small, steeply falling, and highly uncertain~\cite{Romero-Wolf:2017xqe, AlvesBatista:2018zui, Heinze:2019jou, Fang:2013vla, Padovani:2015mba, Fang:2017zjf, Muzio:2019leu, Rodrigues:2020pli, Muzio:2021zud}, though a nonzero flux is guaranteed by the measured UHE cosmic-ray flux. Detecting UHE neutrinos is motivated by important astrophysics questions such as the origin and composition of UHE cosmic rays~\cite{Fang:2016hop, Takami:2007pp, Ahlers:2009rf, Kotera:2010yn, Ahlers:2012rz, Baerwald:2014zga, Aloisio:2015ega, Heinze:2015hhp, Moller:2018isk, vanVliet:2019nse}; however, the flux is so far undetected~\cite{IceCube:2016umi, IceCube:2018fhm, ARA:2019wcf, ANITA:2019wyx, PierreAuger:2019ens}, so we do not yet know which detectors will be optimal.  \Cref{fig:experiments} shows that a wide variety of approaches are proposed~\cite{IceCube:2002eys, BAIKAL:1997iok, KM3Net:2016zxf, ARA:2014fyf, ANITA:2008mzi, ARIANNA:2014fsk, Nam:2016cib, GRAND:2018iaj, Otte:2018uxj, RNO-G:2020rmc, PUEO:2020bnn, POEMMA:2020ykm, Wissel:2020sec, Romero-Wolf:2020pzh, P-ONE:2020ljt, RadarEchoTelescope:2021rca}. While there are encouraging prospects for measuring cross sections for specific assumed fluxes and specific large detectors~\cite{Denton:2020jft, Huang:2021mki, Valera:2022ylt}, it is not known how general these results are.  What are the minimal detector requirements and the statistics needed to make good measurements of the neutrino-nucleon cross sections at the highest energies?

In this paper, we assess these challenges, guided by three principles.  First, instead of considering specific detectors, we focus on the \emph{required} detector properties.  Second, we aim for \emph{model independence} in terms of the assumed neutrino fluxes, theoretical calculations of their propagation in Earth, and detector properties.  Third, we stress the importance of detector \emph{complementarity}, noting that collective measurements over many detectors and energy ranges can be combined.  Bottom line, we show that $\sigma$ can be measured in the UHE range without prior knowledge about the flux, that presently allowed novel-physics scenarios can be tested even with low statistics, and that this can happen relatively soon.

\begin{figure}[b]
    \centering
    \includegraphics[width=0.95\columnwidth]{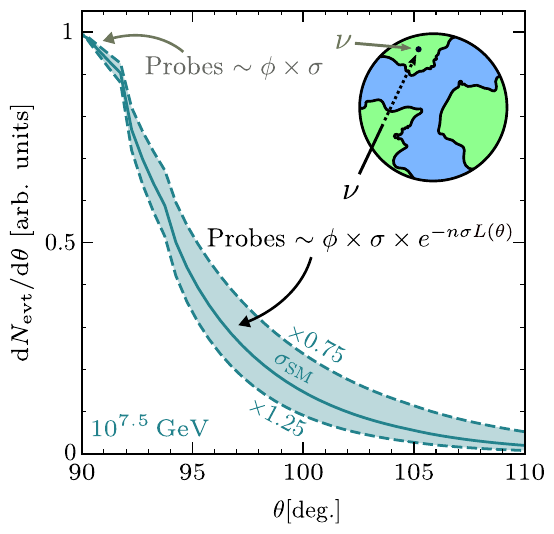}
    \caption{Expected angular profile for a neutrino energy $E_\nu = 10^{7.5} \, \mathrm{GeV}$ and different cross sections.  \emph{Angle-dependent Earth attenuation allows measurement of the cross section}.}
    \label{fig:survival_1}
\end{figure}

\begin{figure}[b]
    \centering
    \includegraphics[width=0.95\columnwidth]{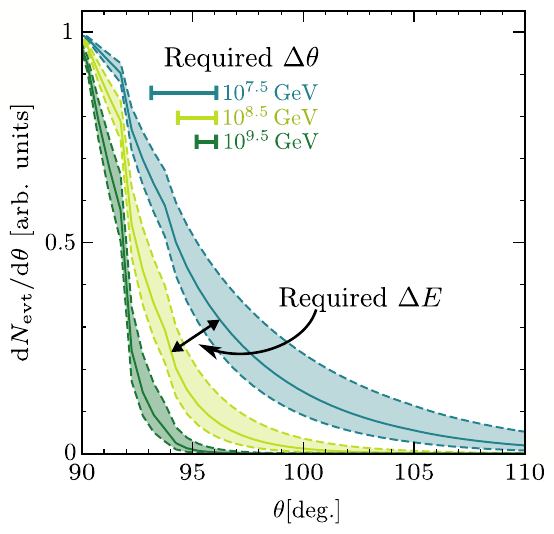}
    \caption{Similar to \cref{fig:survival_1}, but for different energies.  \emph{Discriminating cross sections requires achieving benchmark angular and energy resolutions,} which we show approximately.}
    \label{fig:survival_2}
\end{figure}

The remainder of this paper is organized as follows. In \cref{sec:general_requirements}, we calculate the effects of attenuation and show how these relate to the general requirements for measuring the cross section.  In \cref{sec:sensitivities}, we show how to characterize and compare UHE detector responses, independent of their operating technique.  In \cref{sec:compare}, we calculate how detector sensitivity impacts cross-section measurements.  In \cref{sec:conclusions}, we conclude, and in the Appendices, we provide further details.


\section{General requirements to measure the UHE cross section}
\label{sec:general_requirements}

In this section, we calculate the angular profiles due to neutrino attenuation in Earth, and the detector energy and angular resolution required to use these profiles to measure cross sections.  We find that there are benchmark requirements for these resolutions.  Throughout the paper, we assume an incoming flux of $\mathrm{d}\phi/\mathrm{d}E_\nu \propto E_\nu^{-2.5}$ and focus our calculations on neutrino energies around $10^{8.5} \, \mathrm{GeV}$ (we give more details in \cref{sec:sensitivities}), but the behavior is general.  Some plots for additional fluxes, energies, and resolutions are given in \cref{appendix:energies,appendix:flux}.

\begin{figure*}[t]
    \centering
    \includegraphics[width=\columnwidth,valign=t]{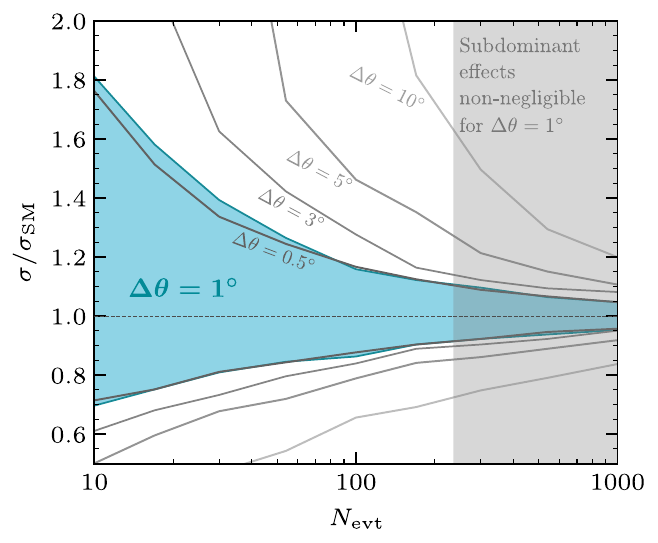}
    \hfill
    \includegraphics[width=\columnwidth,valign=t]{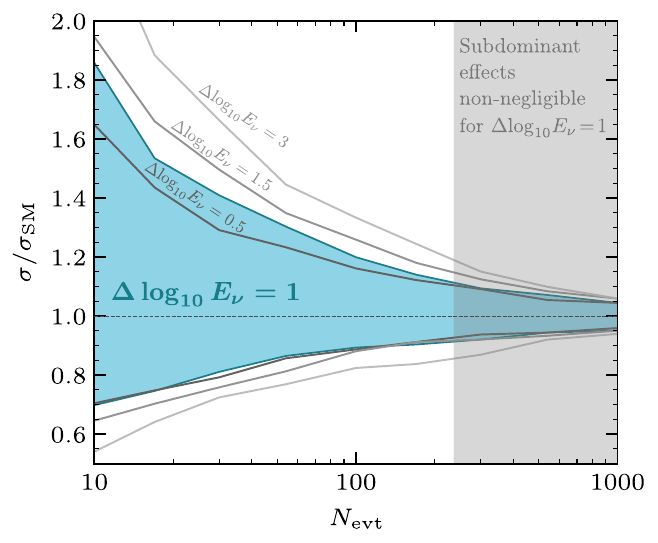}
    \caption{Sensitivity to the cross section for different number of below-horizon events and different angular and energy resolutions.  Here we consider neutrino energies around $10^{8.5} \, \mathrm{GeV}$; the results are similar for other choices as shown in \cref{appendix:energies}.  \emph{Once benchmark angular and energy resolution are reached, the cross-section sensitivity is mostly statistics-limited}.}
    \label{fig:resolution}
\end{figure*}

\Cref{fig:survival_1} shows the physics behind measuring the cross section using Earth attenuation. The relative fluxes of neutrinos from different directions depend only on their trajectories through Earth, as the incoming neutrino flux is expected to be consistent with an isotropic diffuse background.  Flux measurements above or near the horizon probe $\phi \sigma$ while those below it probe $\phi \sigma e^{-n \sigma L(\theta)}$, with $L(\theta)$ the chord length of the neutrino path through Earth. Measuring the angular profile thus allows to probe both unknowns: the flux and cross section. Here $\theta$ is the neutrino arrival zenith angle, measured with respect to the vertical at the point where the neutrino trajectory would exit Earth; and we define ``horizon'' as $\theta=90^\circ$ regardless of the elevation of the detector. For the matter distribution inside Earth, we assume the PREM density profile~\cite{dziewonski1981preliminary}.  We place the detector on top of a 3-km layer of ice, as many of the proposed detectors are on top of, or embedded within, large ice sheets (see \cref{fig:experiments}).  As shown in \cref{appendix:ice}, our results are insensitive to reasonable variations in these choices, especially the latter.  

The angular profile has a strong dependence on the cross section because Earth attenuation is significant.  At a given zenith angle, if even one event is observed, the cross section cannot have been too large (a Poisson expectation of near-zero events cannot fluctuate to one observed event).  Generally, we anticipate precise measurements even with few events because the exponential factor, $e^{-n \sigma L(\theta)}$, is much less than unity.

\Cref{fig:survival_2} gives a first indication that measuring $\sigma$ in the UHE range requires resolving benchmark angular and energy resolutions that are set by the physics of Earth attenuation.  The detector must have sufficient angular resolution to measure the shape of the angular profile.  This is harder here than in the TeV--PeV range, where $\sigma$ is smaller and the angular profile varies slower with $\theta$.  The detector must also have sufficient energy resolution to discriminate if a modified profile is due to a difference in  $\sigma$ or energy.  This is easier here than in the TeV--PeV range, where the variation of  $\sigma$ with energy is stronger.

For our full calculation, we fit for $\sigma/\sigma_{\textrm{SM}}$, the ratio between the cross section and its Standard Model value, using unbinned likelihood as we detail in \cref{appendix:fit}.  In summary, we start with an isotropic power-law flux, $\mathrm{d}\phi/\mathrm{d}E_\nu \propto E_\nu^{-2.5}$. As we detail in \cref{appendix:flux}, a power law is generic for each energy range we consider, and our results are insensitive to the spectral index.  We then add neutrino absorption by Earth, computing $\sigma_\mathrm{SM}$ following Ref.~\cite{Gandhi:1995tf} with the proton parton distribution functions from Ref.~\cite{AbdulKhalek:2022fyi}.  We randomly draw $N_\mathrm{evt}$ events from the 2-D event distribution in $E_\nu$ and $\theta$ and take into account detector resolution as well as the detector efficiency as a function of energy (see the next section). We assume, based on the properties of current and proposed detectors~\cite{ANITA:2008mzi, ARA:2014fyf, Valera:2022ylt}, uniform efficiency as a function of angle in the narrow below-horizon range where all events are expected (see \cref{fig:survival_1,fig:survival_2}); in \cref{sec:compare}, we compare detectors with different above-horizon angular efficiencies. We then fit for $\sigma$, marginalizing over the flux normalization and spectral index. The procedure is repeated many times to obtain the median $1\sigma$ sensitivity.

We neglect \emph{subdominant} effects in our theoretical calculations of expectations, which we find to be justified because they induce $\lesssim 10\%$ corrections.  These include the uncertainty on $\sigma_\mathrm{SM}$~\cite{AbdulKhalek:2022fyi}, non-DIS cross sections~\cite{Zhou:2019vxt, Zhou:2019frk, Garcia:2020jwr, Zhou:2021xuh, Soto:2021vdc}, and $\nu_{\tau}$ and neutral-current regeneration~\cite{Ritz:1987mh, Nicolaidis:1996qu, Halzen:1998be, Kwiecinski:1998yf, Beacom:2001xn, Dutta:2002zc, Arguelles:2022aum} (regeneration is subdominant because the flux is steeply falling; see \cref{appendix:tau} for further discussion and caveats). They only become non-negligible in the very high statistics limit, which may only be obtained in the far future; furthermore, regeneration effects are model-dependent if novel contributions to the cross section are considered. We also neglect backgrounds because they are expected to be negligible relative to the number of events needed to make a measurement of $\sigma$ (see, e.g., Refs.~\cite{PierreAuger:2019ens, ARA:2022rwq, ANITA:2019wyx}). Successful astrophysics measurements also require low backgrounds. Finally, systematic errors are not included in our main results because these mostly become relevant for high statistics and affect different detectors in different ways. For completeness, we discuss systematic effects on the arrival-direction reconstruction in \cref{appendix:fit}.

We next quantify the cross-section sensitivity and the importance of the detector angular and energy resolution. We use three main parameters: the detector resolution in neutrino arrival zenith angle $\Delta\theta$ and in neutrino energy $\Delta \log_{10} E_\nu$, plus the number of detected events below the horizon $N_{\textrm{evt}}$. 

\Cref{fig:resolution} shows that there are benchmark angular and energy resolutions beyond which improvement does not significantly help.  Importantly, achieving these resolutions is challenging but realistic (see, e.g., Refs.~\cite{PUEO:2020bnn, Wissel:2020sec, ARA:2014fyf}).  The cross-section uncertainties are asymmetric because the event distribution depends non-linearly on $\sigma$. We also show in grey the region where the precision on $\sigma$ is better than 10\% and the subdominant effects mentioned above become non-negligible.

The requirement of a benchmark angular resolution in \cref{fig:resolution} (left panel) can be understood from \cref{fig:survival_2}. For $E_\nu = 10^{8.5}$\,GeV as an example, the angular scale that separates negligible from significant attenuation is of order $1^\circ$.  If the detector can resolve this scale, measuring $\sigma$ basically reduces to a counting experiment of events below and above $\sim 91^\circ$; better angular resolution does not significantly improve the measurement. Because the angular profile gets narrower as the neutrino energy increases, the benchmark angular resolution gets somewhat more stringent at higher energies (see \cref{appendix:energies}). 
\Cref{fig:resolution} (right panel) shows that benchmark energy resolution is even easier to meet. For a steeply falling flux, the majority of events for a given detector sensitivity will be detected within a small range in energy. Therefore resolving the energy is less critical than resolving arrival angle; the convolution of a steeply falling flux with a detector threshold is in some sense a built-in energy resolution.

\begin{figure}[t]
\centering
    \includegraphics[width=\columnwidth]{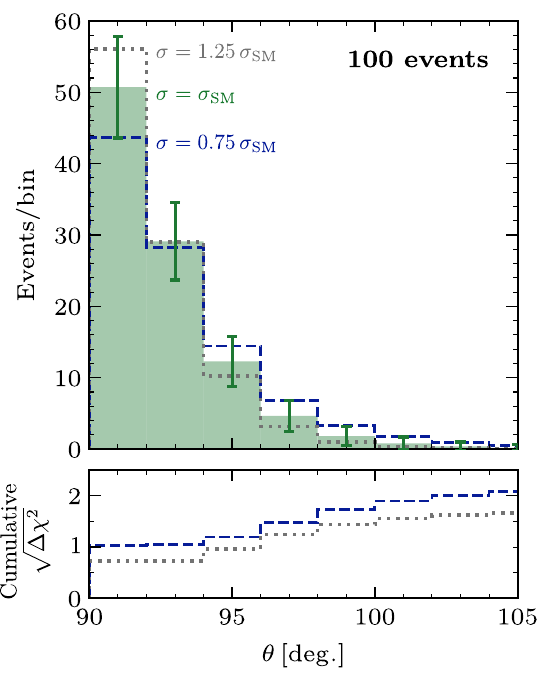} 
    \caption{Simplified illustration of our analysis. The bottom panel shows, after including each bin, the number of sigmas at which modified cross sections are excluded. \emph{Statistics and resolving the angular scale where $\sigma$ affects the shape are key.}}
    \label{fig:evt_distribution}
\end{figure}

\Cref{fig:evt_distribution} shows the impact of statistics, with a simplified illustration (where we generate data once, bin it, and marginalize over energy) of our analysis procedure (where we do not).  We include angular smearing with $\Delta \theta = 1^\circ$.  As we focus on the number of events, all histograms are normalized to 100 events, hence the differences at low zenith angles. The bottom panel shows the cumulative Poissonian $\sqrt{\Delta \chi^2}$ after including data from each bin. The significance accumulates over a wide range of angles, and so the measurement is not dominated by the extreme tail. This figure also illustrates that measuring $\sigma$ boils down to 1) resolving the angular scale where the shape is affected by $\sigma$ and 2) accumulating statistics.

Bringing this all together, \cref{fig:sensitivity_xsec} shows the sensitivity to $\sigma$ at different neutrino energy bins for the benchmark resolutions $\Delta \theta = 1^\circ$ and $\Delta \log_{10} E_\nu = 1$. The black line corresponds to the SM prediction.  At UHE energies, the sensitivity can be better than at TeV--PeV energies (IceCube data) due to the stronger attenuation. We show in \cref{appendix:energies} how the figure changes with varying angular resolution. \textit{Complementary measurements by several UHE detectors, sensitive in different energy ranges, could build up the required statistics over a wide range of $\sqrt{s}$.}

These measurements would be powerful probes of physics at energies beyond collider reach.  We show two novel-physics scenarios in \cref{fig:sensitivity_xsec}: large extra dimensions at a scale of $7 \, \mathrm{TeV}$ and $\sim s^2$ growth of $\sigma$ beyond that scale, computed following Ref.~\cite{Jain:2000pu}; and a leptoquark with mass of $500 \, \mathrm{GeV}$ and coupling of 1 to $\nu_\tau$, $c$ and $s$ quarks, computed following Ref.~\cite{Becirevic:2018uab} (according to Ref.~\cite{Huang:2021mki}, this coupling texture evades LHC limits).  Importantly, these would produce large effects, so high-precision measurements are not required. Physically, this is because leptoquarks would be resonantly produced, and because large extra dimensions entail the exchange of a spin-2 mediator and cross sections grow as $\sim E_\nu^3$~\cite{Jain:2000pu}. Increasing the scales of these novel physics scenarios would produce comparable effects at higher neutrino energies. We also show QCD saturation effects, computed following Ref.~\cite{Arguelles:2015wba}. Even these are in reach if detectors can collectively obtain $\simeq 100$ events above neutrino energies of ${10^9 \, \mathrm{GeV}}$.

Overall, we find that UHE neutrino detectors that can resolve the neutrino direction to better than a few degrees can measure $\sigma$ to better than $\sim$ $^{+65}_{-30}$\% with tens of neutrinos. This is a large number of events, but is achievable with the breadth of proposed and in-development UHE detectors in the literature. Encouragingly, these requirements are not stronger than those necessary to meet the astrophysics goals of UHE detectors. Even with poor resolution in energy --- three decades, for example --- a measurement of $\sigma$ is still robust, although good energy resolution and determining the energy scale are of course necessary for measuring $\sigma$ as a function of energy.

\begin{figure*}[hbtp]
    \centering
    \includegraphics[width=0.329\textwidth]{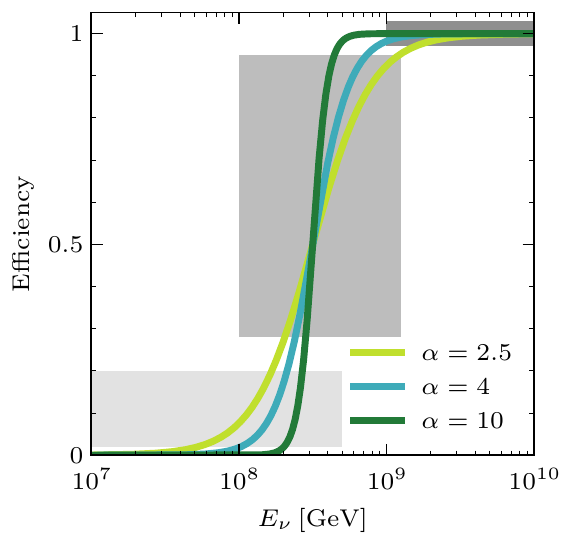} \includegraphics[width=0.329\textwidth]{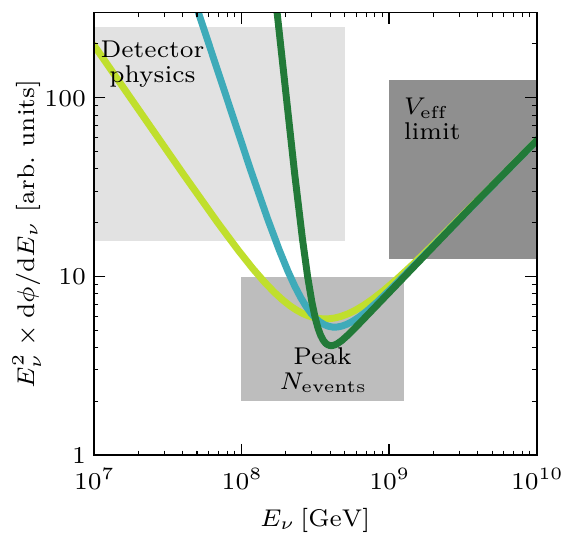} \includegraphics[width=0.329\textwidth]{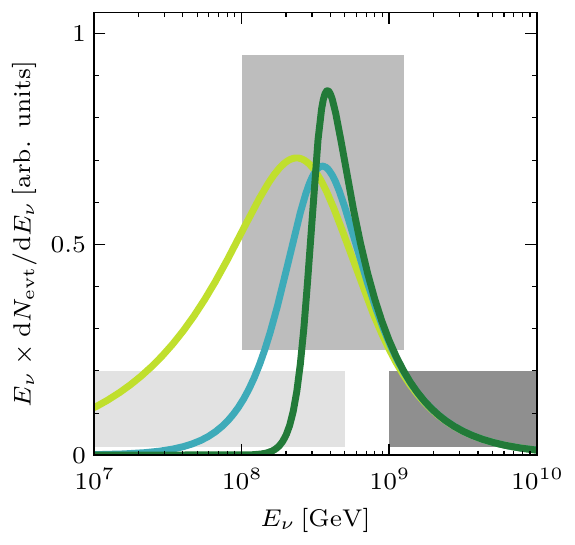}
    \caption{Neutrino detection efficiency (left), flux sensitivity (center), and expected event distribution (right); for different efficiency-growth parameters $\alpha$ and $E_0 = 10^{8.5} \, \mathrm{GeV}$. \emph{Detector physics determines the low-energy behavior, the peak number of events occurs when the efficiency is $\sim 50\%$, and at high energies the detector is limited by effective volume and the number of events is small.}}
    \label{fig:sens}
\end{figure*}

\section{Characterizing UHE detectors}
\label{sec:sensitivities}

In this section, we show how a generic characterization of detector efficiency, which makes it easy to compare different detectors, leads to insights on UHE neutrino detection and cross-section measurements.

UHE neutrino detectors do not directly measure neutrinos, but rather only secondary or even tertiary products of neutrino-induced showers.  Because of the small neutrino interaction probability, these detectors must monitor large volumes of natural material.  A very wide variety of techniques can be used, ranging from passive optical and radio observations to active radar searches.  Nevertheless, we can generally characterize a detector by the efficiency of its response as a function of neutrino energy. Given the steep neutrino spectrum and the slow step-function-like detector efficiency, a response is generally centered at some energy and has some spread around that energy, set by both physical and geometrical factors.  

\Cref{fig:sens} (left panel) shows a general way to describe the detection efficiency as a function of neutrino energy. Similar to the approach in Ref.~\cite{vanSanten:2022wss}, we parameterize the efficiency with a logistic function,
\begin{equation}
    \varepsilon(E_\nu) = \frac{\tanh\left[\alpha\, \log_{10}E_\nu/E_0\right]+1}{2} \,,
    \label{eq:logistic}
\end{equation}
where $E_0$ is the neutrino energy at which the efficiency is 50\% and $\alpha$ characterizes the shape of the rise.  For simplicity, we set the high-energy efficiency to unity, because relatively few events are expected to come from the region above $E_0$ as shown below.  In \cref{sec:general_requirements} above, we use $E_0 = 10^{8.5} \, \mathrm{GeV}$ and $\alpha=4$, a reasonable shape for next-generation experiments~\cite{Ackermann:2022rqc}.

\Cref{fig:sens} (center panel) shows the corresponding minimum flux that can be detected, in units of $E_\nu^2 \, d\phi/dE_\nu = 2.3^{-1} \, E_\nu \, d\phi/d\log_{10}{E_\nu}$. This model-independent differential sensitivity~\cite{Kravchenko:2006qc} is obtained by $E_\nu^2 \mathrm{d}\phi/\mathrm{d}E_\nu \propto E_\nu / [\mathrm{Efficiency}(E_\nu) \times \sigma(E_\nu) \times \mathrm{att}(E_\nu)]$, where the last term is the angle-averaged attenuation. This representation has three distinct and interesting regions that are highlighted in the figure (with corresponding regions indicated in the other panels).  The shape of the curve at low energies is directly set by $\alpha$, that is, by the physics of the detector and what the response is as a function of energy. The minimum of the sensitivity curves is the energy at which the greatest number of neutrinos is expected; it therefore represents a detector's {\it peak sensitivity}. This point is not immediately evident from the efficiency curves in \cref{fig:sens}, but becomes more evident when looking at the right panel. Finally, at high energies the efficiency of the detectors saturates and the sensitivity is determined by the so-called effective volume $V_\mathrm{eff}$, the volume of material to which a detector is sensitive (in the UHE regime this can be far larger than the instrumented volume). Coupled with a falling flux, this results in fewer detected events and decreased sensitivity at the highest energies, even though the efficiency (left panel) is at maximum.

\Cref{fig:sens} (right panel) shows the spectra of detectable events in each case.  Here we show $E_\nu \, dN_\mathrm{evt}/dE_\nu = 2.3^{-1} \, dN_\mathrm{evt}/d\log_{10}{E_\nu}$, which is the number-weighted event rate per energy decade, obtained by multiplying the flux, cross section, efficiency, and angle-averaged attenuation (we only include below-horizon events). For simplicity, we do not include smearing induced by energy resolution. Due to the steeply falling flux, the peaks of the event distributions correspond to the minima of the sensitivity curves (center panel), and detectors with even slightly more efficiency at lower energies see more events: a detector with $\alpha=2.5$ would see $\sim 2$ times more events than one with $\alpha=10$. We show below that, despite the variety of responses, diverse detectors can make robust measurements of $\sigma$ with modest statistics.


\section{Comparison of UHE detector sensitivities}
\label{sec:compare}

\begin{figure*}[t]
    \centering
    \includegraphics[width=\columnwidth, valign=t]{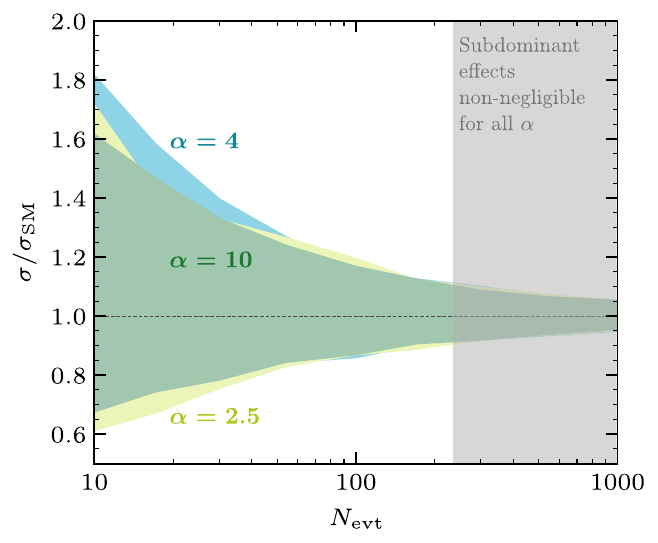} \hfill \includegraphics[width=\columnwidth, valign=t]{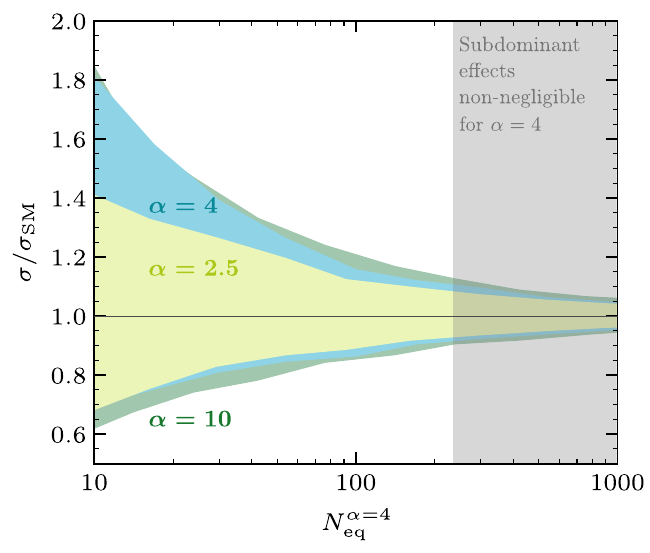}
    \caption{Sensitivity to the cross section for different efficiency-growth parameters, $\alpha$. \textbf{Left panel:} We generate the same number of events for all efficiencies. \textbf{Right panel:} We assume the same flux and scale the number of events relative to the $\alpha=4$ curve. \emph{For the same statistics and resolutions, the sensitivity is nearly the same across experiments}.}
    \label{fig:alphanevents}
\end{figure*}

In the previous section, we explored how to characterize and compare different detectors. Here we apply these results to quantify the impact on cross-section measurements.  We also examine the impact of angular aperture. \\ In all cases, we assume the benchmark resolutions $\Delta \theta = 1^\circ$ and $\Delta \log_{10} E_\nu = 1$.

\Cref{fig:alphanevents} (left panel) shows that, for the same number of events, the \emph{shape} of the sensitivity curve does not significantly affect the constraining power on $\sigma$. This is not a surprise as the bulk of the events fall within the peak sensitivity, see \cref{fig:sens} (right panel). However, as shown there, for a fixed flux the observed number of events $N_\mathrm{evt}$ depends on $\alpha$, so the sensitivities in \cref{fig:alphanevents} (left panel) correspond to different flux \emph{normalizations}.

\Cref{fig:alphanevents} (right panel) shows what happens if we instead keep the normalization of the neutrino flux the same, so that different $\alpha$ correspond to different numbers of events. The horizontal axis shows the number of events that a detector with $\alpha = 4$ would observe, that we denote as $N_\mathrm{eq}^{\alpha=4}$.  For the same flux normalization, detectors with smaller $\alpha$ would observe more events (see \cref{fig:sens}).  This affects sensitivity simply through statistics.

Because we focus on the requirements for cross-section measurements, our main results are given in terms of the number of detected events. The connection between event counts and flux  differs between detectors by a factor $\sim 2$--$3$.  The main reason is flavor, as different detectors are sensitive to different flavors.  Another reason is the inelasticity of the neutrino interaction within the instrumented volume (i.e., how much of the neutrino energy goes into the hadronic cascade, and how that cascade and/or the outgoing lepton are detected). This will also impact the energy resolution $\Delta \log_{10} E_\nu$ on a detector-by-detector basis. Our study moves beyond these detector-specific effects, as well as specific astrophysical fluxes, to understand the problem from a global perspective.

\begin{figure*}[hbtp]
\centering
    \includegraphics[width=\columnwidth, valign=t]{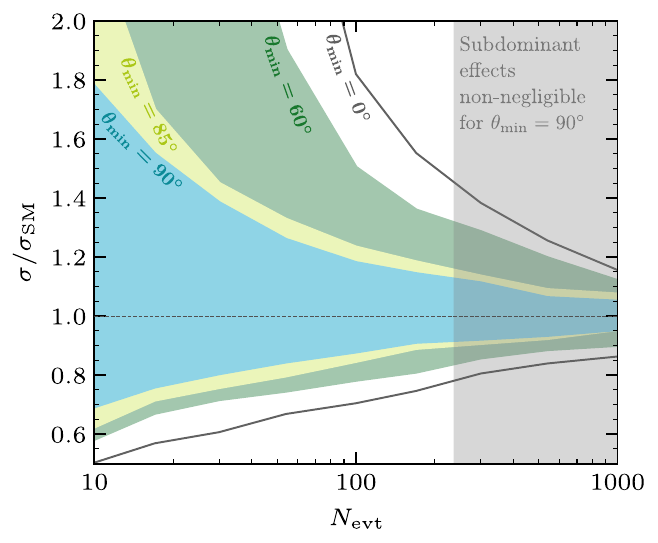} \hfill    \includegraphics[width=\columnwidth, valign=t]{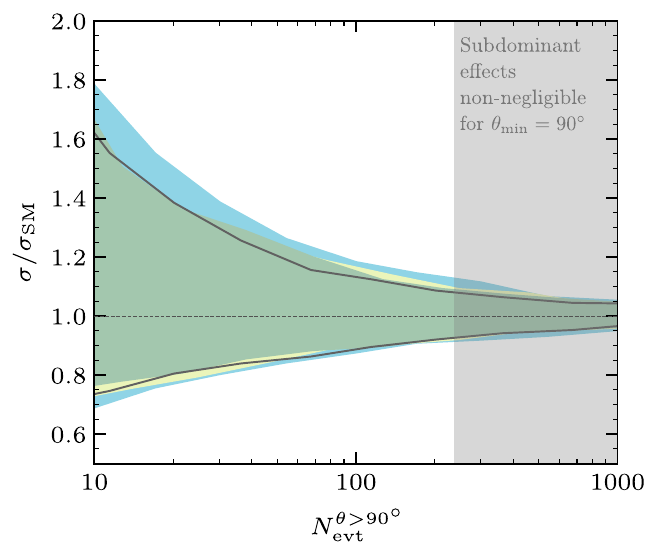}
    \caption{Sensitivity to the cross section for different angular apertures. \textbf{Left panel:} We generate the same number of events for all apertures. \textbf{Right panel:} We scale each curve to have the same number of below-horizon events $N_\mathrm{evt}^{\theta > 90^\circ}$. \emph{The sensitivity is dominated by the below-horizon statistics}.}
    \label{fig:tmin}
\end{figure*}

We next investigate the impact of angular aperture. Some detectors, like $\nu_{\tau}$ detectors on the top of mountains or neutrino detectors in the air or on the surface of ice, have no sensitivity to neutrinos at angles above the horizon since they lie above their detection medium (see \cref{fig:experiments}). Other detectors, however, like embedded in-ice radio or optical detectors, may have a larger angular sensitivity range~\cite{ARA:2014fyf}. Only below-horizon events carry model-independent information on $\sigma$ (see \cref{fig:survival_1}), but a large above-horizon sample would accurately measure the overall normalization of the flux and could add additional information. 

\Cref{fig:tmin} (left panel) shows how the sensitivity depends on the angular aperture of the detector, that we parameterize in terms of an effective zenith cutoff $\theta = \theta_\mathrm{min}$. Here, $N_\mathrm{evt}$ events are distributed from $\theta = 180^\circ$ down to $\theta = \theta_\mathrm{min}$ (for $\theta < 90^\circ$, the angular distributions in \cref{fig:survival_1,fig:survival_2} are flat in $\cos \theta$). $\theta_\mathrm{min}=\{90^\circ, 85^\circ, 60^\circ, 0^\circ\}$ corresponds to 50\%, 55\%, 75\%, and 100\% solid angle coverage, respectively. For consistency, in all cases we assume that the detector is 1.5 km underground within a 3 km ice layer. We also assume that the efficiency is characterized by $\alpha=4$. As expected, detectors that see all of their events below the horizon have better constraining power on $\sigma$ when the total number of detected events stays the same. Here, a smaller $E_0$ would make the results less dependent on $\theta_\mathrm{min}$ due to the higher number of below-horizon events at low energies.

\Cref{fig:tmin} (right panel) shows what happens if we fix the normalization of the flux, such that all detectors see the same number of below-horizon events. Then the difference is negligible: all the information on $\sigma$ comes from below-horizon events. This also justifies ignoring potentially non-negligible above-horizon backgrounds~\cite{deVries:2015oda}.

Overall, we conclude that once angular resolution is $\lesssim 1^{\circ}$ and energy resolution is $\lesssim 1$ decade, all detectors are approximately equally sensitive to $\sigma$, given the same number of detected below-horizon events. Therefore, the critical parameters separating experimental strategies are their abilities to 1) reach these resolutions and 2) scale their effective volume, and increase statistics in their sensitive energy range. Combining various experiments that reach these goals will allow for robust model-independent measurements in different energy ranges, as \cref{fig:sensitivity_xsec} shows.


\section{Conclusions and ways forward}
\label{sec:conclusions}

UHE neutrinos will open outstanding astrophysics opportunities. The guaranteed flux has triggered high interest and many detector ideas, as shown in \cref{fig:experiments}. Success in the astrophysics goals (understanding the cosmic-ray composition and source evolution, multimessenger detection, etc.) will require tens of events and good angular resolution.

These detections will also open outstanding particle physics opportunities. Neutrino-nucleon cross sections will be probed well above the energy scale of colliders, testing many allowed novel-physics scenarios. To get the most out of the potential large investments, it is important to explore the requirements needed for success and the complementarity among several detectors.

In this paper, we present a simple, generic, and accurate model-independent framework to assess the sensitivity to UHE cross sections. Cross sections robustly shape the angular profile of the neutrino flux, whose angular dependence is wide enough and whose energy dependence is mild enough to be measured with reasonable resolutions. We find that the theoretical treatment of propagation can be handled simply, since for the expected statistics we can neglect subleading contribution to the cross-section, its uncertainty, and Earth regeneration effects. On the experimental side, we parameterize detector response in a generic way independent of the detection technique; and we find that we can neglect the precise shape of the detection efficiency and the above-horizon data. We find that once the angular resolution meets a benchmark of about 1 degree, for reasonable energy resolutions nothing but statistics matters; i.e., once the astrophysics requirements are met physics scenarios can also be tested. 

These results lead to an important point regarding \cref{fig:sensitivity_xsec}: \emph{any generic detectors with statistics will do the job, which means that the results of many experiments, with different energy ranges and other properties, can be combined}. Until accumulated statistics exceed the point where subleading effects matter, the potential of each individual detector immediately follows from our results without the need of dedicated studies. 

Our framework sets the stage for further UHE studies. The combined power of several detectors with a few events at different energies should be explored, as this might be the status in the near future. We also provide a tool to fairly compare detectors and understand design choices. Our framework can be immediately generalized to other physics studies, such as distinguishing source models. In addition, for very large statistics measuring neutrino attenuation as a function of angle may shed light on the density profile of the Earth crust. Finally, at even higher energies neutrinos interact in the atmosphere, leading to new observables with different dependencies on $\sigma$~\cite{Kusenko:2001gj, Palomares-Ruiz:2005npx} and different detector requirements that could be systematically explored. Regarding the far future, should a novel physics signal be observed, there will be plenty of opportunities. Our framework allows for deviations from the SM to be identified in a model-independent way. Once identified, the deviations can be explored on a model-by-model basis, ideally in complementarity with future colliders. For high statistics, Earth regeneration effects could be an extra handle on novel physics as they depend on the microphysics details of the underlying interaction model.

In 1930, Pauli~\cite{Pauli:1930pc} proposed the neutrino, later characterizing it as undetectable. By 1934, Bethe and Peierls~\cite{Bethe:1934qn} pointed out that it was in principle detectable, but that the mean free path of an MeV neutrino was comparable to a light year of lead.  Now, nearly 100 years later, there are realistic prospects for measuring neutrino cross sections at energies beyond the reach of any human-made collider. Even more astonishing, this measurement will be made using astrophysical neutrinos, taking advantage of the growth of the cross section with energy and detectors that view gigantic volumes. By looking at the most elusive particles at the highest energies, UHE neutrino detectors will probe a range whose full potential is yet to be determined.

\begin{acknowledgements}

We are grateful for helpful comments from Luis Anchordoqui, Carlos Arguelles, Francis Halzen, Alejandro Ibarra, and especially Amy Connolly, Peter Denton, Krijn de Vries, Kaeli Hughes, Matthew Kirk, Sergio Palomares-Ruiz, Ibrahim Safa, David Saltzberg, and Victor Valera. The work of S.P.\ was supported by National Science Foundation grant No.\ PHY-2012980.  The work of J.F.B.\ was supported by National Science Foundation grant No.\ PHY-2012955. I.E.\ thanks the Instituto de Fisica Teorica (IFT UAM-CSIC) in Madrid for support via the Centro de Excelencia Severo Ochoa Program under Grant CEX2020-001007-S, during the Extended Workshop ``Neutrino Theories,'' where part of this work was developed. Computing resources were provided by the Ohio Supercomputer Center.

\end{acknowledgements}


\newpage
\bibliography{refs}

\begin{thebibliography}{97}%
\makeatletter
\providecommand \@ifxundefined [1]{%
 \@ifx{#1\undefined}
}%
\providecommand \@ifnum [1]{%
 \ifnum #1\expandafter \@firstoftwo
 \else \expandafter \@secondoftwo
 \fi
}%
\providecommand \@ifx [1]{%
 \ifx #1\expandafter \@firstoftwo
 \else \expandafter \@secondoftwo
 \fi
}%
\providecommand \natexlab [1]{#1}%
\providecommand \enquote  [1]{``#1''}%
\providecommand \bibnamefont  [1]{#1}%
\providecommand \bibfnamefont [1]{#1}%
\providecommand \citenamefont [1]{#1}%
\providecommand \href@noop [0]{\@secondoftwo}%
\providecommand \href [0]{\begingroup \@sanitize@url \@href}%
\providecommand \@href[1]{\@@startlink{#1}\@@href}%
\providecommand \@@href[1]{\endgroup#1\@@endlink}%
\providecommand \@sanitize@url [0]{\catcode `\\12\catcode `\$12\catcode
  `\&12\catcode `\#12\catcode `\^12\catcode `\_12\catcode `\%12\relax}%
\providecommand \@@startlink[1]{}%
\providecommand \@@endlink[0]{}%
\providecommand \url  [0]{\begingroup\@sanitize@url \@url }%
\providecommand \@url [1]{\endgroup\@href {#1}{\urlprefix }}%
\providecommand \urlprefix  [0]{URL }%
\providecommand \Eprint [0]{\href }%
\providecommand \doibase [0]{http://dx.doi.org/}%
\providecommand \selectlanguage [0]{\@gobble}%
\providecommand \bibinfo  [0]{\@secondoftwo}%
\providecommand \bibfield  [0]{\@secondoftwo}%
\providecommand \translation [1]{[#1]}%
\providecommand \BibitemOpen [0]{}%
\providecommand \bibitemStop [0]{}%
\providecommand \bibitemNoStop [0]{.\EOS\space}%
\providecommand \EOS [0]{\spacefactor3000\relax}%
\providecommand \BibitemShut  [1]{\csname bibitem#1\endcsname}%
\let\auto@bib@innerbib\@empty
\bibitem [{\citenamefont {Kusenko}\ and\ \citenamefont
  {Weiler}(2002)}]{Kusenko:2001gj}%
  \BibitemOpen
  \bibfield  {author} {\bibinfo {author} {\bibfnamefont {Alexander}\
  \bibnamefont {Kusenko}}\ and\ \bibinfo {author} {\bibfnamefont {Thomas~J.}\
  \bibnamefont {Weiler}},\ }\bibfield  {title} {\enquote {\bibinfo {title}
  {{Neutrino cross-sections at high-energies and the future observations of
  ultrahigh-energy cosmic rays}},}\ }\href {\doibase
  10.1103/PhysRevLett.88.161101} {\bibfield  {journal} {\bibinfo  {journal}
  {Phys. Rev. Lett.}\ }\textbf {\bibinfo {volume} {88}},\ \bibinfo {pages}
  {161101} (\bibinfo {year} {2002})},\ \Eprint
  {http://arxiv.org/abs/hep-ph/0106071} {arXiv:hep-ph/0106071} \BibitemShut
  {NoStop}%
\bibitem [{\citenamefont {Anchordoqui}\ \emph {et~al.}(2002)\citenamefont
  {Anchordoqui}, \citenamefont {Feng}, \citenamefont {Goldberg},\ and\
  \citenamefont {Shapere}}]{Anchordoqui:2001cg}%
  \BibitemOpen
  \bibfield  {author} {\bibinfo {author} {\bibfnamefont {Luis~A.}\ \bibnamefont
  {Anchordoqui}}, \bibinfo {author} {\bibfnamefont {Jonathan~L.}\ \bibnamefont
  {Feng}}, \bibinfo {author} {\bibfnamefont {Haim}\ \bibnamefont {Goldberg}}, \
  and\ \bibinfo {author} {\bibfnamefont {Alfred~D.}\ \bibnamefont {Shapere}},\
  }\bibfield  {title} {\enquote {\bibinfo {title} {{Black holes from cosmic
  rays: Probes of extra dimensions and new limits on TeV scale gravity}},}\
  }\href {\doibase 10.1103/PhysRevD.65.124027} {\bibfield  {journal} {\bibinfo
  {journal} {Phys. Rev. D}\ }\textbf {\bibinfo {volume} {65}},\ \bibinfo
  {pages} {124027} (\bibinfo {year} {2002})},\ \Eprint
  {http://arxiv.org/abs/hep-ph/0112247} {arXiv:hep-ph/0112247} \BibitemShut
  {NoStop}%
\bibitem [{\citenamefont {Hooper}(2002)}]{Hooper:2002yq}%
  \BibitemOpen
  \bibfield  {author} {\bibinfo {author} {\bibfnamefont {Dan}\ \bibnamefont
  {Hooper}},\ }\bibfield  {title} {\enquote {\bibinfo {title} {{Measuring
  high-energy neutrino nucleon cross-sections with future neutrino
  telescopes}},}\ }\href {\doibase 10.1103/PhysRevD.65.097303} {\bibfield
  {journal} {\bibinfo  {journal} {Phys. Rev. D}\ }\textbf {\bibinfo {volume}
  {65}},\ \bibinfo {pages} {097303} (\bibinfo {year} {2002})},\ \Eprint
  {http://arxiv.org/abs/hep-ph/0203239} {arXiv:hep-ph/0203239} \BibitemShut
  {NoStop}%
\bibitem [{\citenamefont {Hussain}\ \emph {et~al.}(2006)\citenamefont
  {Hussain}, \citenamefont {Marfatia}, \citenamefont {McKay},\ and\
  \citenamefont {Seckel}}]{Hussain:2006wg}%
  \BibitemOpen
  \bibfield  {author} {\bibinfo {author} {\bibfnamefont {S.}~\bibnamefont
  {Hussain}}, \bibinfo {author} {\bibfnamefont {D.}~\bibnamefont {Marfatia}},
  \bibinfo {author} {\bibfnamefont {D.~W.}\ \bibnamefont {McKay}}, \ and\
  \bibinfo {author} {\bibfnamefont {D.}~\bibnamefont {Seckel}},\ }\bibfield
  {title} {\enquote {\bibinfo {title} {{Cross section dependence of event rates
  at neutrino telescopes}},}\ }\href {\doibase 10.1103/PhysRevLett.97.161101}
  {\bibfield  {journal} {\bibinfo  {journal} {Phys. Rev. Lett.}\ }\textbf
  {\bibinfo {volume} {97}},\ \bibinfo {pages} {161101} (\bibinfo {year}
  {2006})},\ \Eprint {http://arxiv.org/abs/hep-ph/0606246}
  {arXiv:hep-ph/0606246} \BibitemShut {NoStop}%
\bibitem [{\citenamefont {Borriello}\ \emph {et~al.}(2008)\citenamefont
  {Borriello}, \citenamefont {Cuoco}, \citenamefont {Mangano}, \citenamefont
  {Miele}, \citenamefont {Pastor}, \citenamefont {Pisanti},\ and\ \citenamefont
  {Serpico}}]{Borriello:2007cs}%
  \BibitemOpen
  \bibfield  {author} {\bibinfo {author} {\bibfnamefont {E.}~\bibnamefont
  {Borriello}}, \bibinfo {author} {\bibfnamefont {A.}~\bibnamefont {Cuoco}},
  \bibinfo {author} {\bibfnamefont {G.}~\bibnamefont {Mangano}}, \bibinfo
  {author} {\bibfnamefont {G.}~\bibnamefont {Miele}}, \bibinfo {author}
  {\bibfnamefont {S.}~\bibnamefont {Pastor}}, \bibinfo {author} {\bibfnamefont
  {O.}~\bibnamefont {Pisanti}}, \ and\ \bibinfo {author} {\bibfnamefont
  {P.~D.}\ \bibnamefont {Serpico}},\ }\bibfield  {title} {\enquote {\bibinfo
  {title} {{Disentangling neutrino-nucleon cross section and high energy
  neutrino flux with a km$^3$ neutrino telescope}},}\ }\href {\doibase
  10.1103/PhysRevD.77.045019} {\bibfield  {journal} {\bibinfo  {journal} {Phys.
  Rev. D}\ }\textbf {\bibinfo {volume} {77}},\ \bibinfo {pages} {045019}
  (\bibinfo {year} {2008})},\ \Eprint {http://arxiv.org/abs/0711.0152}
  {arXiv:0711.0152 [astro-ph]} \BibitemShut {NoStop}%
\bibitem [{\citenamefont {Hussain}\ \emph {et~al.}(2008)\citenamefont
  {Hussain}, \citenamefont {Marfatia},\ and\ \citenamefont
  {McKay}}]{Hussain:2007ba}%
  \BibitemOpen
  \bibfield  {author} {\bibinfo {author} {\bibfnamefont {S.}~\bibnamefont
  {Hussain}}, \bibinfo {author} {\bibfnamefont {D.}~\bibnamefont {Marfatia}}, \
  and\ \bibinfo {author} {\bibfnamefont {D.~W.}\ \bibnamefont {McKay}},\
  }\bibfield  {title} {\enquote {\bibinfo {title} {{Upward shower rates at
  neutrino telescopes directly determine the neutrino flux}},}\ }\href
  {\doibase 10.1103/PhysRevD.77.107304} {\bibfield  {journal} {\bibinfo
  {journal} {Phys. Rev. D}\ }\textbf {\bibinfo {volume} {77}},\ \bibinfo
  {pages} {107304} (\bibinfo {year} {2008})},\ \Eprint
  {http://arxiv.org/abs/0711.4374} {arXiv:0711.4374 [hep-ph]} \BibitemShut
  {NoStop}%
\bibitem [{\citenamefont {Connolly}\ \emph {et~al.}(2011)\citenamefont
  {Connolly}, \citenamefont {Thorne},\ and\ \citenamefont
  {Waters}}]{Connolly:2011vc}%
  \BibitemOpen
  \bibfield  {author} {\bibinfo {author} {\bibfnamefont {Amy}\ \bibnamefont
  {Connolly}}, \bibinfo {author} {\bibfnamefont {Robert~S.}\ \bibnamefont
  {Thorne}}, \ and\ \bibinfo {author} {\bibfnamefont {David}\ \bibnamefont
  {Waters}},\ }\bibfield  {title} {\enquote {\bibinfo {title} {{Calculation of
  High Energy Neutrino-Nucleon Cross Sections and Uncertainties Using the MSTW
  Parton Distribution Functions and Implications for Future Experiments}},}\
  }\href {\doibase 10.1103/PhysRevD.83.113009} {\bibfield  {journal} {\bibinfo
  {journal} {Phys. Rev. D}\ }\textbf {\bibinfo {volume} {83}},\ \bibinfo
  {pages} {113009} (\bibinfo {year} {2011})},\ \Eprint
  {http://arxiv.org/abs/1102.0691} {arXiv:1102.0691 [hep-ph]} \BibitemShut
  {NoStop}%
\bibitem [{\citenamefont {Marfatia}\ \emph {et~al.}(2015)\citenamefont
  {Marfatia}, \citenamefont {McKay},\ and\ \citenamefont
  {Weiler}}]{Marfatia:2015hva}%
  \BibitemOpen
  \bibfield  {author} {\bibinfo {author} {\bibfnamefont {D.}~\bibnamefont
  {Marfatia}}, \bibinfo {author} {\bibfnamefont {D.~W.}\ \bibnamefont {McKay}},
  \ and\ \bibinfo {author} {\bibfnamefont {T.~J.}\ \bibnamefont {Weiler}},\
  }\bibfield  {title} {\enquote {\bibinfo {title} {{New physics with
  ultra-high-energy neutrinos}},}\ }\href {\doibase
  10.1016/j.physletb.2015.07.002} {\bibfield  {journal} {\bibinfo  {journal}
  {Phys. Lett. B}\ }\textbf {\bibinfo {volume} {748}},\ \bibinfo {pages}
  {113--116} (\bibinfo {year} {2015})},\ \Eprint
  {http://arxiv.org/abs/1502.06337} {arXiv:1502.06337 [hep-ph]} \BibitemShut
  {NoStop}%
\bibitem [{\citenamefont {Bustamante}\ and\ \citenamefont
  {Connolly}(2019)}]{Bustamante:2017xuy}%
  \BibitemOpen
  \bibfield  {author} {\bibinfo {author} {\bibfnamefont {Mauricio}\
  \bibnamefont {Bustamante}}\ and\ \bibinfo {author} {\bibfnamefont {Amy}\
  \bibnamefont {Connolly}},\ }\bibfield  {title} {\enquote {\bibinfo {title}
  {{Extracting the Energy-Dependent Neutrino-Nucleon Cross Section above 10 TeV
  Using IceCube Showers}},}\ }\href {\doibase 10.1103/PhysRevLett.122.041101}
  {\bibfield  {journal} {\bibinfo  {journal} {Phys. Rev. Lett.}\ }\textbf
  {\bibinfo {volume} {122}},\ \bibinfo {pages} {041101} (\bibinfo {year}
  {2019})},\ \Eprint {http://arxiv.org/abs/1711.11043} {arXiv:1711.11043
  [astro-ph.HE]} \BibitemShut {NoStop}%
\bibitem [{\citenamefont {Abbasi}\ \emph {et~al.}(2020)\citenamefont {Abbasi}
  \emph {et~al.}}]{IceCube:2020rnc}%
  \BibitemOpen
  \bibfield  {author} {\bibinfo {author} {\bibfnamefont {R.}~\bibnamefont
  {Abbasi}} \emph {et~al.} (\bibinfo {collaboration} {IceCube}),\ }\bibfield
  {title} {\enquote {\bibinfo {title} {{Measurement of the high-energy
  all-flavor neutrino-nucleon cross section with IceCube}},}\ }\href {\doibase
  10.1103/PhysRevD.104.022001} {\bibfield  {journal} {\bibinfo  {journal}
  {Phys. Rev. D}\ } (\bibinfo {year} {2020}),\ 10.1103/PhysRevD.104.022001},\
  \Eprint {http://arxiv.org/abs/2011.03560} {arXiv:2011.03560 [hep-ex]}
  \BibitemShut {NoStop}%
\bibitem [{\citenamefont {Romero-Wolf}\ and\ \citenamefont
  {Ave}(2018)}]{Romero-Wolf:2017xqe}%
  \BibitemOpen
  \bibfield  {author} {\bibinfo {author} {\bibfnamefont {Andr\'es}\
  \bibnamefont {Romero-Wolf}}\ and\ \bibinfo {author} {\bibfnamefont
  {M\'aximo}\ \bibnamefont {Ave}},\ }\bibfield  {title} {\enquote {\bibinfo
  {title} {{Bayesian Inference Constraints on Astrophysical Production of
  Ultra-high Energy Cosmic Rays and Cosmogenic Neutrino Flux Predictions}},}\
  }\href {\doibase 10.1088/1475-7516/2018/07/025} {\bibfield  {journal}
  {\bibinfo  {journal} {JCAP}\ }\textbf {\bibinfo {volume} {07}},\ \bibinfo
  {pages} {025} (\bibinfo {year} {2018})},\ \Eprint
  {http://arxiv.org/abs/1712.07290} {arXiv:1712.07290 [astro-ph.HE]}
  \BibitemShut {NoStop}%
\bibitem [{\citenamefont {Alves~Batista}\ \emph {et~al.}(2019)\citenamefont
  {Alves~Batista}, \citenamefont {de~Almeida}, \citenamefont {Lago},\ and\
  \citenamefont {Kotera}}]{AlvesBatista:2018zui}%
  \BibitemOpen
  \bibfield  {author} {\bibinfo {author} {\bibfnamefont {Rafael}\ \bibnamefont
  {Alves~Batista}}, \bibinfo {author} {\bibfnamefont {Rogerio~M.}\ \bibnamefont
  {de~Almeida}}, \bibinfo {author} {\bibfnamefont {Bruno}\ \bibnamefont
  {Lago}}, \ and\ \bibinfo {author} {\bibfnamefont {Kumiko}\ \bibnamefont
  {Kotera}},\ }\bibfield  {title} {\enquote {\bibinfo {title} {{Cosmogenic
  photon and neutrino fluxes in the Auger era}},}\ }\href {\doibase
  10.1088/1475-7516/2019/01/002} {\bibfield  {journal} {\bibinfo  {journal}
  {JCAP}\ }\textbf {\bibinfo {volume} {01}},\ \bibinfo {pages} {002} (\bibinfo
  {year} {2019})},\ \Eprint {http://arxiv.org/abs/1806.10879} {arXiv:1806.10879
  [astro-ph.HE]} \BibitemShut {NoStop}%
\bibitem [{\citenamefont {Heinze}\ \emph {et~al.}(2019)\citenamefont {Heinze},
  \citenamefont {Fedynitch}, \citenamefont {Boncioli},\ and\ \citenamefont
  {Winter}}]{Heinze:2019jou}%
  \BibitemOpen
  \bibfield  {author} {\bibinfo {author} {\bibfnamefont {Jonas}\ \bibnamefont
  {Heinze}}, \bibinfo {author} {\bibfnamefont {Anatoli}\ \bibnamefont
  {Fedynitch}}, \bibinfo {author} {\bibfnamefont {Denise}\ \bibnamefont
  {Boncioli}}, \ and\ \bibinfo {author} {\bibfnamefont {Walter}\ \bibnamefont
  {Winter}},\ }\bibfield  {title} {\enquote {\bibinfo {title} {{A new view on
  Auger data and cosmogenic neutrinos in light of different nuclear
  disintegration and air-shower models}},}\ }\href {\doibase
  10.3847/1538-4357/ab05ce} {\bibfield  {journal} {\bibinfo  {journal}
  {Astrophys. J.}\ }\textbf {\bibinfo {volume} {873}},\ \bibinfo {pages} {88}
  (\bibinfo {year} {2019})},\ \Eprint {http://arxiv.org/abs/1901.03338}
  {arXiv:1901.03338 [astro-ph.HE]} \BibitemShut {NoStop}%
\bibitem [{\citenamefont {Fang}\ \emph {et~al.}(2014)\citenamefont {Fang},
  \citenamefont {Kotera}, \citenamefont {Murase},\ and\ \citenamefont
  {Olinto}}]{Fang:2013vla}%
  \BibitemOpen
  \bibfield  {author} {\bibinfo {author} {\bibfnamefont {Ke}~\bibnamefont
  {Fang}}, \bibinfo {author} {\bibfnamefont {Kumiko}\ \bibnamefont {Kotera}},
  \bibinfo {author} {\bibfnamefont {Kohta}\ \bibnamefont {Murase}}, \ and\
  \bibinfo {author} {\bibfnamefont {Angela~V.}\ \bibnamefont {Olinto}},\
  }\bibfield  {title} {\enquote {\bibinfo {title} {{Testing the Newborn Pulsar
  Origin of Ultrahigh Energy Cosmic Rays with EeV Neutrinos}},}\ }\href
  {\doibase 10.1103/PhysRevD.90.103005} {\bibfield  {journal} {\bibinfo
  {journal} {Phys. Rev. D}\ }\textbf {\bibinfo {volume} {90}},\ \bibinfo
  {pages} {103005} (\bibinfo {year} {2014})},\ \bibinfo {note} {[Erratum:
  Phys.Rev.D 92, 129901(E) (2015)]},\ \Eprint {http://arxiv.org/abs/1311.2044}
  {arXiv:1311.2044 [astro-ph.HE]} \BibitemShut {NoStop}%
\bibitem [{\citenamefont {Padovani}\ \emph {et~al.}(2015)\citenamefont
  {Padovani}, \citenamefont {Petropoulou}, \citenamefont {Giommi},\ and\
  \citenamefont {Resconi}}]{Padovani:2015mba}%
  \BibitemOpen
  \bibfield  {author} {\bibinfo {author} {\bibfnamefont {P.}~\bibnamefont
  {Padovani}}, \bibinfo {author} {\bibfnamefont {M.}~\bibnamefont
  {Petropoulou}}, \bibinfo {author} {\bibfnamefont {P.}~\bibnamefont {Giommi}},
  \ and\ \bibinfo {author} {\bibfnamefont {E.}~\bibnamefont {Resconi}},\
  }\bibfield  {title} {\enquote {\bibinfo {title} {{A simplified view of
  blazars: the neutrino background}},}\ }\href {\doibase 10.1093/mnras/stv1467}
  {\bibfield  {journal} {\bibinfo  {journal} {Mon. Not. Roy. Astron. Soc.}\
  }\textbf {\bibinfo {volume} {452}},\ \bibinfo {pages} {1877--1887} (\bibinfo
  {year} {2015})},\ \Eprint {http://arxiv.org/abs/1506.09135} {arXiv:1506.09135
  [astro-ph.HE]} \BibitemShut {NoStop}%
\bibitem [{\citenamefont {Fang}\ and\ \citenamefont
  {Murase}(2018)}]{Fang:2017zjf}%
  \BibitemOpen
  \bibfield  {author} {\bibinfo {author} {\bibfnamefont {Ke}~\bibnamefont
  {Fang}}\ and\ \bibinfo {author} {\bibfnamefont {Kohta}\ \bibnamefont
  {Murase}},\ }\bibfield  {title} {\enquote {\bibinfo {title} {{Linking
  High-Energy Cosmic Particles by Black Hole Jets Embedded in Large-Scale
  Structures}},}\ }\href {\doibase 10.1038/s41567-017-0025-4} {\bibfield
  {journal} {\bibinfo  {journal} {Nature Phys.}\ }\textbf {\bibinfo {volume}
  {14}},\ \bibinfo {pages} {396--398} (\bibinfo {year} {2018})},\ \Eprint
  {http://arxiv.org/abs/1704.00015} {arXiv:1704.00015 [astro-ph.HE]}
  \BibitemShut {NoStop}%
\bibitem [{\citenamefont {Muzio}\ \emph {et~al.}(2019)\citenamefont {Muzio},
  \citenamefont {Unger},\ and\ \citenamefont {Farrar}}]{Muzio:2019leu}%
  \BibitemOpen
  \bibfield  {author} {\bibinfo {author} {\bibfnamefont {Marco~Stein}\
  \bibnamefont {Muzio}}, \bibinfo {author} {\bibfnamefont {Michael}\
  \bibnamefont {Unger}}, \ and\ \bibinfo {author} {\bibfnamefont {Glennys~R.}\
  \bibnamefont {Farrar}},\ }\bibfield  {title} {\enquote {\bibinfo {title}
  {{Progress towards characterizing ultrahigh energy cosmic ray sources}},}\
  }\href {\doibase 10.1103/PhysRevD.100.103008} {\bibfield  {journal} {\bibinfo
   {journal} {Phys. Rev. D}\ }\textbf {\bibinfo {volume} {100}},\ \bibinfo
  {pages} {103008} (\bibinfo {year} {2019})},\ \Eprint
  {http://arxiv.org/abs/1906.06233} {arXiv:1906.06233 [astro-ph.HE]}
  \BibitemShut {NoStop}%
\bibitem [{\citenamefont {Rodrigues}\ \emph {et~al.}(2021)\citenamefont
  {Rodrigues}, \citenamefont {Heinze}, \citenamefont {Palladino}, \citenamefont
  {van Vliet},\ and\ \citenamefont {Winter}}]{Rodrigues:2020pli}%
  \BibitemOpen
  \bibfield  {author} {\bibinfo {author} {\bibfnamefont {Xavier}\ \bibnamefont
  {Rodrigues}}, \bibinfo {author} {\bibfnamefont {Jonas}\ \bibnamefont
  {Heinze}}, \bibinfo {author} {\bibfnamefont {Andrea}\ \bibnamefont
  {Palladino}}, \bibinfo {author} {\bibfnamefont {Arjen}\ \bibnamefont {van
  Vliet}}, \ and\ \bibinfo {author} {\bibfnamefont {Walter}\ \bibnamefont
  {Winter}},\ }\bibfield  {title} {\enquote {\bibinfo {title} {{Active Galactic
  Nuclei Jets as the Origin of Ultrahigh-Energy Cosmic Rays and Perspectives
  for the Detection of Astrophysical Source Neutrinos at EeV Energies}},}\
  }\href {\doibase 10.1103/PhysRevLett.126.191101} {\bibfield  {journal}
  {\bibinfo  {journal} {Phys. Rev. Lett.}\ }\textbf {\bibinfo {volume} {126}},\
  \bibinfo {pages} {191101} (\bibinfo {year} {2021})},\ \Eprint
  {http://arxiv.org/abs/2003.08392} {arXiv:2003.08392 [astro-ph.HE]}
  \BibitemShut {NoStop}%
\bibitem [{\citenamefont {Muzio}\ \emph {et~al.}(2022)\citenamefont {Muzio},
  \citenamefont {Farrar},\ and\ \citenamefont {Unger}}]{Muzio:2021zud}%
  \BibitemOpen
  \bibfield  {author} {\bibinfo {author} {\bibfnamefont {Marco~Stein}\
  \bibnamefont {Muzio}}, \bibinfo {author} {\bibfnamefont {Glennys~R.}\
  \bibnamefont {Farrar}}, \ and\ \bibinfo {author} {\bibfnamefont {Michael}\
  \bibnamefont {Unger}},\ }\bibfield  {title} {\enquote {\bibinfo {title}
  {{Probing the environments surrounding ultrahigh energy cosmic ray
  accelerators and their implications for astrophysical neutrinos}},}\ }\href
  {\doibase 10.1103/PhysRevD.105.023022} {\bibfield  {journal} {\bibinfo
  {journal} {Phys. Rev. D}\ }\textbf {\bibinfo {volume} {105}},\ \bibinfo
  {pages} {023022} (\bibinfo {year} {2022})},\ \Eprint
  {http://arxiv.org/abs/2108.05512} {arXiv:2108.05512 [astro-ph.HE]}
  \BibitemShut {NoStop}%
\bibitem [{\citenamefont {Fang}\ \emph {et~al.}(2016)\citenamefont {Fang},
  \citenamefont {Kotera}, \citenamefont {Miller}, \citenamefont {Murase},\ and\
  \citenamefont {Oikonomou}}]{Fang:2016hop}%
  \BibitemOpen
  \bibfield  {author} {\bibinfo {author} {\bibfnamefont {Ke}~\bibnamefont
  {Fang}}, \bibinfo {author} {\bibfnamefont {Kumiko}\ \bibnamefont {Kotera}},
  \bibinfo {author} {\bibfnamefont {M.~Coleman}\ \bibnamefont {Miller}},
  \bibinfo {author} {\bibfnamefont {Kohta}\ \bibnamefont {Murase}}, \ and\
  \bibinfo {author} {\bibfnamefont {Foteini}\ \bibnamefont {Oikonomou}},\
  }\bibfield  {title} {\enquote {\bibinfo {title} {{Identifying
  Ultrahigh-Energy Cosmic-Ray Accelerators with Future Ultrahigh-Energy
  Neutrino Detectors}},}\ }\href {\doibase 10.1088/1475-7516/2016/12/017}
  {\bibfield  {journal} {\bibinfo  {journal} {JCAP}\ }\textbf {\bibinfo
  {volume} {12}},\ \bibinfo {pages} {017} (\bibinfo {year} {2016})},\ \Eprint
  {http://arxiv.org/abs/1609.08027} {arXiv:1609.08027 [astro-ph.HE]}
  \BibitemShut {NoStop}%
\bibitem [{\citenamefont {Takami}\ \emph {et~al.}(2009)\citenamefont {Takami},
  \citenamefont {Murase}, \citenamefont {Nagataki},\ and\ \citenamefont
  {Sato}}]{Takami:2007pp}%
  \BibitemOpen
  \bibfield  {author} {\bibinfo {author} {\bibfnamefont {Hajime}\ \bibnamefont
  {Takami}}, \bibinfo {author} {\bibfnamefont {Kohta}\ \bibnamefont {Murase}},
  \bibinfo {author} {\bibfnamefont {Shigehiro}\ \bibnamefont {Nagataki}}, \
  and\ \bibinfo {author} {\bibfnamefont {Katsuhiko}\ \bibnamefont {Sato}},\
  }\bibfield  {title} {\enquote {\bibinfo {title} {{Cosmogenic neutrinos as a
  probe of the transition from Galactic to extragalactic cosmic rays}},}\
  }\href {\doibase 10.1016/j.astropartphys.2009.01.006} {\bibfield  {journal}
  {\bibinfo  {journal} {Astropart. Phys.}\ }\textbf {\bibinfo {volume} {31}},\
  \bibinfo {pages} {201--211} (\bibinfo {year} {2009})},\ \Eprint
  {http://arxiv.org/abs/0704.0979} {arXiv:0704.0979 [astro-ph]} \BibitemShut
  {NoStop}%
\bibitem [{\citenamefont {Ahlers}\ \emph {et~al.}(2009)\citenamefont {Ahlers},
  \citenamefont {Anchordoqui},\ and\ \citenamefont {Sarkar}}]{Ahlers:2009rf}%
  \BibitemOpen
  \bibfield  {author} {\bibinfo {author} {\bibfnamefont {Markus}\ \bibnamefont
  {Ahlers}}, \bibinfo {author} {\bibfnamefont {Luis~A.}\ \bibnamefont
  {Anchordoqui}}, \ and\ \bibinfo {author} {\bibfnamefont {Subir}\ \bibnamefont
  {Sarkar}},\ }\bibfield  {title} {\enquote {\bibinfo {title} {{Neutrino
  diagnostics of ultra-high energy cosmic ray protons}},}\ }\href {\doibase
  10.1103/PhysRevD.79.083009} {\bibfield  {journal} {\bibinfo  {journal} {Phys.
  Rev. D}\ }\textbf {\bibinfo {volume} {79}},\ \bibinfo {pages} {083009}
  (\bibinfo {year} {2009})},\ \Eprint {http://arxiv.org/abs/0902.3993}
  {arXiv:0902.3993 [astro-ph.HE]} \BibitemShut {NoStop}%
\bibitem [{\citenamefont {Kotera}\ \emph {et~al.}(2010)\citenamefont {Kotera},
  \citenamefont {Allard},\ and\ \citenamefont {Olinto}}]{Kotera:2010yn}%
  \BibitemOpen
  \bibfield  {author} {\bibinfo {author} {\bibfnamefont {K.}~\bibnamefont
  {Kotera}}, \bibinfo {author} {\bibfnamefont {D.}~\bibnamefont {Allard}}, \
  and\ \bibinfo {author} {\bibfnamefont {A.~V.}\ \bibnamefont {Olinto}},\
  }\bibfield  {title} {\enquote {\bibinfo {title} {{Cosmogenic Neutrinos:
  parameter space and detectabilty from PeV to ZeV}},}\ }\href {\doibase
  10.1088/1475-7516/2010/10/013} {\bibfield  {journal} {\bibinfo  {journal}
  {JCAP}\ }\textbf {\bibinfo {volume} {10}},\ \bibinfo {pages} {013} (\bibinfo
  {year} {2010})},\ \Eprint {http://arxiv.org/abs/1009.1382} {arXiv:1009.1382
  [astro-ph.HE]} \BibitemShut {NoStop}%
\bibitem [{\citenamefont {Ahlers}\ and\ \citenamefont
  {Halzen}(2012)}]{Ahlers:2012rz}%
  \BibitemOpen
  \bibfield  {author} {\bibinfo {author} {\bibfnamefont {Markus}\ \bibnamefont
  {Ahlers}}\ and\ \bibinfo {author} {\bibfnamefont {Francis}\ \bibnamefont
  {Halzen}},\ }\bibfield  {title} {\enquote {\bibinfo {title} {{Minimal
  Cosmogenic Neutrinos}},}\ }\href {\doibase 10.1103/PhysRevD.86.083010}
  {\bibfield  {journal} {\bibinfo  {journal} {Phys. Rev. D}\ }\textbf {\bibinfo
  {volume} {86}},\ \bibinfo {pages} {083010} (\bibinfo {year} {2012})},\
  \Eprint {http://arxiv.org/abs/1208.4181} {arXiv:1208.4181 [astro-ph.HE]}
  \BibitemShut {NoStop}%
\bibitem [{\citenamefont {Baerwald}\ \emph {et~al.}(2015)\citenamefont
  {Baerwald}, \citenamefont {Bustamante},\ and\ \citenamefont
  {Winter}}]{Baerwald:2014zga}%
  \BibitemOpen
  \bibfield  {author} {\bibinfo {author} {\bibfnamefont {Philipp}\ \bibnamefont
  {Baerwald}}, \bibinfo {author} {\bibfnamefont {Mauricio}\ \bibnamefont
  {Bustamante}}, \ and\ \bibinfo {author} {\bibfnamefont {Walter}\ \bibnamefont
  {Winter}},\ }\bibfield  {title} {\enquote {\bibinfo {title} {{Are gamma-ray
  bursts the sources of ultra-high energy cosmic rays?}}}\ }\href {\doibase
  10.1016/j.astropartphys.2014.07.007} {\bibfield  {journal} {\bibinfo
  {journal} {Astropart. Phys.}\ }\textbf {\bibinfo {volume} {62}},\ \bibinfo
  {pages} {66--91} (\bibinfo {year} {2015})},\ \Eprint
  {http://arxiv.org/abs/1401.1820} {arXiv:1401.1820 [astro-ph.HE]} \BibitemShut
  {NoStop}%
\bibitem [{\citenamefont {Aloisio}\ \emph {et~al.}(2015)\citenamefont
  {Aloisio}, \citenamefont {Boncioli}, \citenamefont {di~Matteo}, \citenamefont
  {Grillo}, \citenamefont {Petrera},\ and\ \citenamefont
  {Salamida}}]{Aloisio:2015ega}%
  \BibitemOpen
  \bibfield  {author} {\bibinfo {author} {\bibfnamefont {R.}~\bibnamefont
  {Aloisio}}, \bibinfo {author} {\bibfnamefont {D.}~\bibnamefont {Boncioli}},
  \bibinfo {author} {\bibfnamefont {A}~\bibnamefont {di~Matteo}}, \bibinfo
  {author} {\bibfnamefont {A.~F.}\ \bibnamefont {Grillo}}, \bibinfo {author}
  {\bibfnamefont {S.}~\bibnamefont {Petrera}}, \ and\ \bibinfo {author}
  {\bibfnamefont {F.}~\bibnamefont {Salamida}},\ }\bibfield  {title} {\enquote
  {\bibinfo {title} {{Cosmogenic neutrinos and ultra-high energy cosmic ray
  models}},}\ }\href {\doibase 10.1088/1475-7516/2015/10/006} {\bibfield
  {journal} {\bibinfo  {journal} {JCAP}\ }\textbf {\bibinfo {volume} {10}},\
  \bibinfo {pages} {006} (\bibinfo {year} {2015})},\ \Eprint
  {http://arxiv.org/abs/1505.04020} {arXiv:1505.04020 [astro-ph.HE]}
  \BibitemShut {NoStop}%
\bibitem [{\citenamefont {Heinze}\ \emph {et~al.}(2016)\citenamefont {Heinze},
  \citenamefont {Boncioli}, \citenamefont {Bustamante},\ and\ \citenamefont
  {Winter}}]{Heinze:2015hhp}%
  \BibitemOpen
  \bibfield  {author} {\bibinfo {author} {\bibfnamefont {Jonas}\ \bibnamefont
  {Heinze}}, \bibinfo {author} {\bibfnamefont {Denise}\ \bibnamefont
  {Boncioli}}, \bibinfo {author} {\bibfnamefont {Mauricio}\ \bibnamefont
  {Bustamante}}, \ and\ \bibinfo {author} {\bibfnamefont {Walter}\ \bibnamefont
  {Winter}},\ }\bibfield  {title} {\enquote {\bibinfo {title} {{Cosmogenic
  Neutrinos Challenge the Cosmic Ray Proton Dip Model}},}\ }\href {\doibase
  10.3847/0004-637X/825/2/122} {\bibfield  {journal} {\bibinfo  {journal}
  {Astrophys. J.}\ }\textbf {\bibinfo {volume} {825}},\ \bibinfo {pages} {122}
  (\bibinfo {year} {2016})},\ \Eprint {http://arxiv.org/abs/1512.05988}
  {arXiv:1512.05988 [astro-ph.HE]} \BibitemShut {NoStop}%
\bibitem [{\citenamefont {M\o{}ller}\ \emph {et~al.}(2019)\citenamefont
  {M\o{}ller}, \citenamefont {Denton},\ and\ \citenamefont
  {Tamborra}}]{Moller:2018isk}%
  \BibitemOpen
  \bibfield  {author} {\bibinfo {author} {\bibfnamefont {Klaes}\ \bibnamefont
  {M\o{}ller}}, \bibinfo {author} {\bibfnamefont {Peter~B.}\ \bibnamefont
  {Denton}}, \ and\ \bibinfo {author} {\bibfnamefont {Irene}\ \bibnamefont
  {Tamborra}},\ }\bibfield  {title} {\enquote {\bibinfo {title} {{Cosmogenic
  Neutrinos Through the GRAND Lens Unveil the Nature of Cosmic
  Accelerators}},}\ }\href {\doibase 10.1088/1475-7516/2019/05/047} {\bibfield
  {journal} {\bibinfo  {journal} {JCAP}\ }\textbf {\bibinfo {volume} {05}},\
  \bibinfo {pages} {047} (\bibinfo {year} {2019})},\ \Eprint
  {http://arxiv.org/abs/1809.04866} {arXiv:1809.04866 [astro-ph.HE]}
  \BibitemShut {NoStop}%
\bibitem [{\citenamefont {van Vliet}\ \emph {et~al.}(2019)\citenamefont {van
  Vliet}, \citenamefont {Alves~Batista},\ and\ \citenamefont
  {H\"orandel}}]{vanVliet:2019nse}%
  \BibitemOpen
  \bibfield  {author} {\bibinfo {author} {\bibfnamefont {Arjen}\ \bibnamefont
  {van Vliet}}, \bibinfo {author} {\bibfnamefont {Rafael}\ \bibnamefont
  {Alves~Batista}}, \ and\ \bibinfo {author} {\bibfnamefont {J\"org~R.}\
  \bibnamefont {H\"orandel}},\ }\bibfield  {title} {\enquote {\bibinfo {title}
  {{Determining the fraction of cosmic-ray protons at ultrahigh energies with
  cosmogenic neutrinos}},}\ }\href {\doibase 10.1103/PhysRevD.100.021302}
  {\bibfield  {journal} {\bibinfo  {journal} {Phys. Rev. D}\ }\textbf {\bibinfo
  {volume} {100}},\ \bibinfo {pages} {021302(R)} (\bibinfo {year} {2019})},\
  \Eprint {http://arxiv.org/abs/1901.01899} {arXiv:1901.01899 [astro-ph.HE]}
  \BibitemShut {NoStop}%
\bibitem [{\citenamefont {Aartsen}\ \emph {et~al.}(2016)\citenamefont {Aartsen}
  \emph {et~al.}}]{IceCube:2016umi}%
  \BibitemOpen
  \bibfield  {author} {\bibinfo {author} {\bibfnamefont {M.~G.}\ \bibnamefont
  {Aartsen}} \emph {et~al.} (\bibinfo {collaboration} {IceCube}),\ }\bibfield
  {title} {\enquote {\bibinfo {title} {{Observation and Characterization of a
  Cosmic Muon Neutrino Flux from the Northern Hemisphere using six years of
  IceCube data}},}\ }\href {\doibase 10.3847/0004-637X/833/1/3} {\bibfield
  {journal} {\bibinfo  {journal} {Astrophys. J.}\ }\textbf {\bibinfo {volume}
  {833}},\ \bibinfo {pages} {3} (\bibinfo {year} {2016})},\ \Eprint
  {http://arxiv.org/abs/1607.08006} {arXiv:1607.08006 [astro-ph.HE]}
  \BibitemShut {NoStop}%
\bibitem [{\citenamefont {Aartsen}\ \emph {et~al.}(2018)\citenamefont {Aartsen}
  \emph {et~al.}}]{IceCube:2018fhm}%
  \BibitemOpen
  \bibfield  {author} {\bibinfo {author} {\bibfnamefont {M.~G.}\ \bibnamefont
  {Aartsen}} \emph {et~al.} (\bibinfo {collaboration} {IceCube}),\ }\bibfield
  {title} {\enquote {\bibinfo {title} {{Differential limit on the
  extremely-high-energy cosmic neutrino flux in the presence of astrophysical
  background from nine years of IceCube data}},}\ }\href {\doibase
  10.1103/PhysRevD.98.062003} {\bibfield  {journal} {\bibinfo  {journal} {Phys.
  Rev. D}\ }\textbf {\bibinfo {volume} {98}},\ \bibinfo {pages} {062003}
  (\bibinfo {year} {2018})},\ \Eprint {http://arxiv.org/abs/1807.01820}
  {arXiv:1807.01820 [astro-ph.HE]} \BibitemShut {NoStop}%
\bibitem [{\citenamefont {Allison}\ \emph
  {et~al.}(2020{\natexlab{a}})\citenamefont {Allison} \emph
  {et~al.}}]{ARA:2019wcf}%
  \BibitemOpen
  \bibfield  {author} {\bibinfo {author} {\bibfnamefont {P.}~\bibnamefont
  {Allison}} \emph {et~al.} (\bibinfo {collaboration} {ARA}),\ }\bibfield
  {title} {\enquote {\bibinfo {title} {{Constraints on the diffuse flux of
  ultrahigh energy neutrinos from four years of Askaryan Radio Array data in
  two stations}},}\ }\href {\doibase 10.1103/PhysRevD.102.043021} {\bibfield
  {journal} {\bibinfo  {journal} {Phys. Rev. D}\ }\textbf {\bibinfo {volume}
  {102}},\ \bibinfo {pages} {043021} (\bibinfo {year} {2020}{\natexlab{a}})},\
  \Eprint {http://arxiv.org/abs/1912.00987} {arXiv:1912.00987 [astro-ph.HE]}
  \BibitemShut {NoStop}%
\bibitem [{\citenamefont {Gorham}\ \emph {et~al.}(2019)\citenamefont {Gorham}
  \emph {et~al.}}]{ANITA:2019wyx}%
  \BibitemOpen
  \bibfield  {author} {\bibinfo {author} {\bibfnamefont {P.~W.}\ \bibnamefont
  {Gorham}} \emph {et~al.} (\bibinfo {collaboration} {ANITA}),\ }\bibfield
  {title} {\enquote {\bibinfo {title} {{Constraints on the ultrahigh-energy
  cosmic neutrino flux from the fourth flight of ANITA}},}\ }\href {\doibase
  10.1103/PhysRevD.99.122001} {\bibfield  {journal} {\bibinfo  {journal} {Phys.
  Rev. D}\ }\textbf {\bibinfo {volume} {99}},\ \bibinfo {pages} {122001}
  (\bibinfo {year} {2019})},\ \Eprint {http://arxiv.org/abs/1902.04005}
  {arXiv:1902.04005 [astro-ph.HE]} \BibitemShut {NoStop}%
\bibitem [{\citenamefont {Aab}\ \emph {et~al.}(2019)\citenamefont {Aab} \emph
  {et~al.}}]{PierreAuger:2019ens}%
  \BibitemOpen
  \bibfield  {author} {\bibinfo {author} {\bibfnamefont {Alexander}\
  \bibnamefont {Aab}} \emph {et~al.} (\bibinfo {collaboration} {Pierre
  Auger}),\ }\bibfield  {title} {\enquote {\bibinfo {title} {{Probing the
  origin of ultra-high-energy cosmic rays with neutrinos in the EeV energy
  range using the Pierre Auger Observatory}},}\ }\href {\doibase
  10.1088/1475-7516/2019/10/022} {\bibfield  {journal} {\bibinfo  {journal}
  {JCAP}\ }\textbf {\bibinfo {volume} {10}},\ \bibinfo {pages} {022} (\bibinfo
  {year} {2019})},\ \Eprint {http://arxiv.org/abs/1906.07422} {arXiv:1906.07422
  [astro-ph.HE]} \BibitemShut {NoStop}%
\bibitem [{\citenamefont {Ahrens}\ \emph {et~al.}(2003)\citenamefont {Ahrens}
  \emph {et~al.}}]{IceCube:2002eys}%
  \BibitemOpen
  \bibfield  {author} {\bibinfo {author} {\bibfnamefont {J.}~\bibnamefont
  {Ahrens}} \emph {et~al.} (\bibinfo {collaboration} {IceCube}),\ }\bibfield
  {title} {\enquote {\bibinfo {title} {{Icecube - the next generation neutrino
  telescope at the south pole}},}\ }\href {\doibase
  10.1016/S0920-5632(03)01337-9} {\bibfield  {journal} {\bibinfo  {journal}
  {Nucl. Phys. B Proc. Suppl.}\ }\textbf {\bibinfo {volume} {118}},\ \bibinfo
  {pages} {388--395} (\bibinfo {year} {2003})},\ \Eprint
  {http://arxiv.org/abs/astro-ph/0209556} {arXiv:astro-ph/0209556} \BibitemShut
  {NoStop}%
\bibitem [{\citenamefont {Belolaptikov}\ \emph {et~al.}(1997)\citenamefont
  {Belolaptikov} \emph {et~al.}}]{BAIKAL:1997iok}%
  \BibitemOpen
  \bibfield  {author} {\bibinfo {author} {\bibfnamefont {I.~A.}\ \bibnamefont
  {Belolaptikov}} \emph {et~al.} (\bibinfo {collaboration} {BAIKAL}),\
  }\bibfield  {title} {\enquote {\bibinfo {title} {{The Baikal underwater
  neutrino telescope: Design, performance and first results}},}\ }\href
  {\doibase 10.1016/S0927-6505(97)00022-4} {\bibfield  {journal} {\bibinfo
  {journal} {Astropart. Phys.}\ }\textbf {\bibinfo {volume} {7}},\ \bibinfo
  {pages} {263--282} (\bibinfo {year} {1997})}\BibitemShut {NoStop}%
\bibitem [{\citenamefont {Adrian-Martinez}\ \emph {et~al.}(2016)\citenamefont
  {Adrian-Martinez} \emph {et~al.}}]{KM3Net:2016zxf}%
  \BibitemOpen
  \bibfield  {author} {\bibinfo {author} {\bibfnamefont {S.}~\bibnamefont
  {Adrian-Martinez}} \emph {et~al.} (\bibinfo {collaboration} {KM3Net}),\
  }\bibfield  {title} {\enquote {\bibinfo {title} {{Letter of intent for KM3NeT
  2.0}},}\ }\href {\doibase 10.1088/0954-3899/43/8/084001} {\bibfield
  {journal} {\bibinfo  {journal} {J. Phys. G}\ }\textbf {\bibinfo {volume}
  {43}},\ \bibinfo {pages} {084001} (\bibinfo {year} {2016})},\ \Eprint
  {http://arxiv.org/abs/1601.07459} {arXiv:1601.07459 [astro-ph.IM]}
  \BibitemShut {NoStop}%
\bibitem [{\citenamefont {Allison}\ \emph {et~al.}(2015)\citenamefont {Allison}
  \emph {et~al.}}]{ARA:2014fyf}%
  \BibitemOpen
  \bibfield  {author} {\bibinfo {author} {\bibfnamefont {P.}~\bibnamefont
  {Allison}} \emph {et~al.} (\bibinfo {collaboration} {ARA}),\ }\bibfield
  {title} {\enquote {\bibinfo {title} {{First Constraints on the Ultra-High
  Energy Neutrino Flux from a Prototype Station of the Askaryan Radio
  Array}},}\ }\href {\doibase 10.1016/j.astropartphys.2015.04.006} {\bibfield
  {journal} {\bibinfo  {journal} {Astropart. Phys.}\ }\textbf {\bibinfo
  {volume} {70}},\ \bibinfo {pages} {62--80} (\bibinfo {year} {2015})},\
  \Eprint {http://arxiv.org/abs/1404.5285} {arXiv:1404.5285 [astro-ph.HE]}
  \BibitemShut {NoStop}%
\bibitem [{\citenamefont {Gorham}\ \emph {et~al.}(2009)\citenamefont {Gorham}
  \emph {et~al.}}]{ANITA:2008mzi}%
  \BibitemOpen
  \bibfield  {author} {\bibinfo {author} {\bibfnamefont {P.~W.}\ \bibnamefont
  {Gorham}} \emph {et~al.} (\bibinfo {collaboration} {ANITA}),\ }\bibfield
  {title} {\enquote {\bibinfo {title} {{The Antarctic Impulsive Transient
  Antenna Ultra-high Energy Neutrino Detector Design, Performance, and
  Sensitivity for 2006-2007 Balloon Flight}},}\ }\href {\doibase
  10.1016/j.astropartphys.2009.05.003} {\bibfield  {journal} {\bibinfo
  {journal} {Astropart. Phys.}\ }\textbf {\bibinfo {volume} {32}},\ \bibinfo
  {pages} {10--41} (\bibinfo {year} {2009})},\ \Eprint
  {http://arxiv.org/abs/0812.1920} {arXiv:0812.1920 [astro-ph]} \BibitemShut
  {NoStop}%
\bibitem [{\citenamefont {Barwick}\ \emph {et~al.}(2015)\citenamefont {Barwick}
  \emph {et~al.}}]{ARIANNA:2014fsk}%
  \BibitemOpen
  \bibfield  {author} {\bibinfo {author} {\bibfnamefont {S.~W.}\ \bibnamefont
  {Barwick}} \emph {et~al.} (\bibinfo {collaboration} {ARIANNA}),\ }\bibfield
  {title} {\enquote {\bibinfo {title} {{A First Search for Cosmogenic Neutrinos
  with the ARIANNA Hexagonal Radio Array}},}\ }\href {\doibase
  10.1016/j.astropartphys.2015.04.002} {\bibfield  {journal} {\bibinfo
  {journal} {Astropart. Phys.}\ }\textbf {\bibinfo {volume} {70}},\ \bibinfo
  {pages} {12--26} (\bibinfo {year} {2015})},\ \Eprint
  {http://arxiv.org/abs/1410.7352} {arXiv:1410.7352 [astro-ph.HE]} \BibitemShut
  {NoStop}%
\bibitem [{\citenamefont {Nam}\ \emph {et~al.}(2016)\citenamefont {Nam} \emph
  {et~al.}}]{Nam:2016cib}%
  \BibitemOpen
  \bibfield  {author} {\bibinfo {author} {\bibfnamefont {J.~W.}\ \bibnamefont
  {Nam}} \emph {et~al.},\ }\bibfield  {title} {\enquote {\bibinfo {title}
  {{Design and implementation of the TAROGE experiment}},}\ }\href {\doibase
  10.1142/S0218271816450139} {\bibfield  {journal} {\bibinfo  {journal} {Int.
  J. Mod. Phys. D}\ }\textbf {\bibinfo {volume} {25}},\ \bibinfo {pages}
  {1645013} (\bibinfo {year} {2016})}\BibitemShut {NoStop}%
\bibitem [{\citenamefont {\'Alvarez-Mu\~niz}\ \emph {et~al.}(2020)\citenamefont
  {\'Alvarez-Mu\~niz} \emph {et~al.}}]{GRAND:2018iaj}%
  \BibitemOpen
  \bibfield  {author} {\bibinfo {author} {\bibfnamefont {Jaime}\ \bibnamefont
  {\'Alvarez-Mu\~niz}} \emph {et~al.} (\bibinfo {collaboration} {GRAND}),\
  }\bibfield  {title} {\enquote {\bibinfo {title} {{The Giant Radio Array for
  Neutrino Detection (GRAND): Science and Design}},}\ }\href {\doibase
  10.1007/s11433-018-9385-7} {\bibfield  {journal} {\bibinfo  {journal} {Sci.
  China Phys. Mech. Astron.}\ }\textbf {\bibinfo {volume} {63}},\ \bibinfo
  {pages} {219501} (\bibinfo {year} {2020})},\ \Eprint
  {http://arxiv.org/abs/1810.09994} {arXiv:1810.09994 [astro-ph.HE]}
  \BibitemShut {NoStop}%
\bibitem [{\citenamefont {Otte}(2019)}]{Otte:2018uxj}%
  \BibitemOpen
  \bibfield  {author} {\bibinfo {author} {\bibfnamefont {Adam~Nepomuk}\
  \bibnamefont {Otte}},\ }\bibfield  {title} {\enquote {\bibinfo {title}
  {{Studies of an air-shower imaging system for the detection of
  ultrahigh-energy neutrinos}},}\ }\href {\doibase 10.1103/PhysRevD.99.083012}
  {\bibfield  {journal} {\bibinfo  {journal} {Phys. Rev. D}\ }\textbf {\bibinfo
  {volume} {99}},\ \bibinfo {pages} {083012} (\bibinfo {year} {2019})},\
  \Eprint {http://arxiv.org/abs/1811.09287} {arXiv:1811.09287 [astro-ph.IM]}
  \BibitemShut {NoStop}%
\bibitem [{\citenamefont {Aguilar}\ \emph {et~al.}(2021)\citenamefont {Aguilar}
  \emph {et~al.}}]{RNO-G:2020rmc}%
  \BibitemOpen
  \bibfield  {author} {\bibinfo {author} {\bibfnamefont {J.~A.}\ \bibnamefont
  {Aguilar}} \emph {et~al.} (\bibinfo {collaboration} {RNO-G}),\ }\bibfield
  {title} {\enquote {\bibinfo {title} {{Design and Sensitivity of the Radio
  Neutrino Observatory in Greenland (RNO-G)}},}\ }\href {\doibase
  10.1088/1748-0221/16/03/P03025} {\bibfield  {journal} {\bibinfo  {journal}
  {JINST}\ }\textbf {\bibinfo {volume} {16}},\ \bibinfo {pages} {P03025}
  (\bibinfo {year} {2021})},\ \Eprint {http://arxiv.org/abs/2010.12279}
  {arXiv:2010.12279 [astro-ph.IM]} \BibitemShut {NoStop}%
\bibitem [{\citenamefont {Abarr}\ \emph {et~al.}(2021)\citenamefont {Abarr}
  \emph {et~al.}}]{PUEO:2020bnn}%
  \BibitemOpen
  \bibfield  {author} {\bibinfo {author} {\bibfnamefont {Q.}~\bibnamefont
  {Abarr}} \emph {et~al.} (\bibinfo {collaboration} {PUEO}),\ }\bibfield
  {title} {\enquote {\bibinfo {title} {{The Payload for Ultrahigh Energy
  Observations (PUEO): a white paper}},}\ }\href {\doibase
  10.1088/1748-0221/16/08/P08035} {\bibfield  {journal} {\bibinfo  {journal}
  {JINST}\ }\textbf {\bibinfo {volume} {16}},\ \bibinfo {pages} {P08035}
  (\bibinfo {year} {2021})},\ \Eprint {http://arxiv.org/abs/2010.02892}
  {arXiv:2010.02892 [astro-ph.IM]} \BibitemShut {NoStop}%
\bibitem [{\citenamefont {Olinto}\ \emph {et~al.}(2021)\citenamefont {Olinto}
  \emph {et~al.}}]{POEMMA:2020ykm}%
  \BibitemOpen
  \bibfield  {author} {\bibinfo {author} {\bibfnamefont {A.~V.}\ \bibnamefont
  {Olinto}} \emph {et~al.} (\bibinfo {collaboration} {POEMMA}),\ }\bibfield
  {title} {\enquote {\bibinfo {title} {{The POEMMA (Probe of Extreme
  Multi-Messenger Astrophysics) observatory}},}\ }\href {\doibase
  10.1088/1475-7516/2021/06/007} {\bibfield  {journal} {\bibinfo  {journal}
  {JCAP}\ }\textbf {\bibinfo {volume} {06}},\ \bibinfo {pages} {007} (\bibinfo
  {year} {2021})},\ \Eprint {http://arxiv.org/abs/2012.07945} {arXiv:2012.07945
  [astro-ph.IM]} \BibitemShut {NoStop}%
\bibitem [{\citenamefont {Wissel}\ \emph {et~al.}(2020)\citenamefont {Wissel}
  \emph {et~al.}}]{Wissel:2020sec}%
  \BibitemOpen
  \bibfield  {author} {\bibinfo {author} {\bibfnamefont {Stephanie}\
  \bibnamefont {Wissel}} \emph {et~al.},\ }\bibfield  {title} {\enquote
  {\bibinfo {title} {{Prospects for high-elevation radio detection of
  \ensuremath{>}100 PeV tau neutrinos}},}\ }\href {\doibase
  10.1088/1475-7516/2020/11/065} {\bibfield  {journal} {\bibinfo  {journal}
  {JCAP}\ }\textbf {\bibinfo {volume} {11}},\ \bibinfo {pages} {065} (\bibinfo
  {year} {2020})},\ \Eprint {http://arxiv.org/abs/2004.12718} {arXiv:2004.12718
  [astro-ph.IM]} \BibitemShut {NoStop}%
\bibitem [{\citenamefont {Romero-Wolf}\ \emph {et~al.}(2020)\citenamefont
  {Romero-Wolf} \emph {et~al.}}]{Romero-Wolf:2020pzh}%
  \BibitemOpen
  \bibfield  {author} {\bibinfo {author} {\bibfnamefont {Andres}\ \bibnamefont
  {Romero-Wolf}} \emph {et~al.},\ }\bibfield  {title} {\enquote {\bibinfo
  {title} {{An Andean Deep-Valley Detector for High-Energy Tau Neutrinos}},}\
  }in\ \href@noop {} {\emph {\bibinfo {booktitle} {{Latin American Strategy
  Forum for Research Infrastructure}}}}\ (\bibinfo {year} {2020})\ \Eprint
  {http://arxiv.org/abs/2002.06475} {arXiv:2002.06475 [astro-ph.IM]}
  \BibitemShut {NoStop}%
\bibitem [{\citenamefont {Agostini}\ \emph {et~al.}(2020)\citenamefont
  {Agostini} \emph {et~al.}}]{P-ONE:2020ljt}%
  \BibitemOpen
  \bibfield  {author} {\bibinfo {author} {\bibfnamefont {Matteo}\ \bibnamefont
  {Agostini}} \emph {et~al.} (\bibinfo {collaboration} {P-ONE}),\ }\bibfield
  {title} {\enquote {\bibinfo {title} {{The Pacific Ocean Neutrino
  Experiment}},}\ }\href {\doibase 10.1038/s41550-020-1182-4} {\bibfield
  {journal} {\bibinfo  {journal} {Nature Astron.}\ }\textbf {\bibinfo {volume}
  {4}},\ \bibinfo {pages} {913--915} (\bibinfo {year} {2020})},\ \Eprint
  {http://arxiv.org/abs/2005.09493} {arXiv:2005.09493 [astro-ph.HE]}
  \BibitemShut {NoStop}%
\bibitem [{\citenamefont {Prohira}\ \emph
  {et~al.}(2021{\natexlab{a}})\citenamefont {Prohira} \emph
  {et~al.}}]{RadarEchoTelescope:2021rca}%
  \BibitemOpen
  \bibfield  {author} {\bibinfo {author} {\bibfnamefont {S.}~\bibnamefont
  {Prohira}} \emph {et~al.} (\bibinfo {collaboration} {Radar Echo Telescope}),\
  }\bibfield  {title} {\enquote {\bibinfo {title} {{The Radar Echo Telescope
  for Cosmic Rays: Pathfinder experiment for a next-generation neutrino
  observatory}},}\ }\href {\doibase 10.1103/PhysRevD.104.102006} {\bibfield
  {journal} {\bibinfo  {journal} {Phys. Rev. D}\ }\textbf {\bibinfo {volume}
  {104}},\ \bibinfo {pages} {102006} (\bibinfo {year} {2021}{\natexlab{a}})},\
  \Eprint {http://arxiv.org/abs/2104.00459} {arXiv:2104.00459 [astro-ph.IM]}
  \BibitemShut {NoStop}%
\bibitem [{\citenamefont {Denton}\ and\ \citenamefont
  {Kini}(2020)}]{Denton:2020jft}%
  \BibitemOpen
  \bibfield  {author} {\bibinfo {author} {\bibfnamefont {Peter~B.}\
  \bibnamefont {Denton}}\ and\ \bibinfo {author} {\bibfnamefont {Yves}\
  \bibnamefont {Kini}},\ }\bibfield  {title} {\enquote {\bibinfo {title}
  {{Ultra-High-Energy Tau Neutrino Cross Sections with GRAND and POEMMA}},}\
  }\href {\doibase 10.1103/PhysRevD.102.123019} {\bibfield  {journal} {\bibinfo
   {journal} {Phys. Rev. D}\ }\textbf {\bibinfo {volume} {102}},\ \bibinfo
  {pages} {123019} (\bibinfo {year} {2020})},\ \Eprint
  {http://arxiv.org/abs/2007.10334} {arXiv:2007.10334 [astro-ph.HE]}
  \BibitemShut {NoStop}%
\bibitem [{\citenamefont {Huang}\ \emph {et~al.}(2022)\citenamefont {Huang},
  \citenamefont {Jana}, \citenamefont {Lindner},\ and\ \citenamefont
  {Rodejohann}}]{Huang:2021mki}%
  \BibitemOpen
  \bibfield  {author} {\bibinfo {author} {\bibfnamefont {Guo-yuan}\
  \bibnamefont {Huang}}, \bibinfo {author} {\bibfnamefont {Sudip}\ \bibnamefont
  {Jana}}, \bibinfo {author} {\bibfnamefont {Manfred}\ \bibnamefont {Lindner}},
  \ and\ \bibinfo {author} {\bibfnamefont {Werner}\ \bibnamefont
  {Rodejohann}},\ }\bibfield  {title} {\enquote {\bibinfo {title} {{Probing new
  physics at future tau neutrino telescopes}},}\ }\href {\doibase
  10.1088/1475-7516/2022/02/038} {\bibfield  {journal} {\bibinfo  {journal}
  {JCAP}\ }\textbf {\bibinfo {volume} {02}},\ \bibinfo {pages} {038} (\bibinfo
  {year} {2022})},\ \Eprint {http://arxiv.org/abs/2112.09476} {arXiv:2112.09476
  [hep-ph]} \BibitemShut {NoStop}%
\bibitem [{\citenamefont {Valera}\ \emph {et~al.}(2022)\citenamefont {Valera},
  \citenamefont {Bustamante},\ and\ \citenamefont {Glaser}}]{Valera:2022ylt}%
  \BibitemOpen
  \bibfield  {author} {\bibinfo {author} {\bibfnamefont {Victor~Branco}\
  \bibnamefont {Valera}}, \bibinfo {author} {\bibfnamefont {Mauricio}\
  \bibnamefont {Bustamante}}, \ and\ \bibinfo {author} {\bibfnamefont
  {Christian}\ \bibnamefont {Glaser}},\ }\bibfield  {title} {\enquote {\bibinfo
  {title} {{The ultra-high-energy neutrino-nucleon cross section: measurement
  forecasts for an era of cosmic EeV-neutrino discovery}},}\ }\href {\doibase
  10.1007/JHEP06(2022)105} {\bibfield  {journal} {\bibinfo  {journal} {JHEP}\
  }\textbf {\bibinfo {volume} {06}},\ \bibinfo {pages} {105} (\bibinfo {year}
  {2022})},\ \Eprint {http://arxiv.org/abs/2204.04237} {arXiv:2204.04237
  [hep-ph]} \BibitemShut {NoStop}%
\bibitem [{\citenamefont {Dziewonski}\ and\ \citenamefont
  {Anderson}(1981)}]{dziewonski1981preliminary}%
  \BibitemOpen
  \bibfield  {author} {\bibinfo {author} {\bibfnamefont {Adam~M}\ \bibnamefont
  {Dziewonski}}\ and\ \bibinfo {author} {\bibfnamefont {Don~L}\ \bibnamefont
  {Anderson}},\ }\bibfield  {title} {\enquote {\bibinfo {title} {Preliminary
  reference earth model},}\ }\href@noop {} {\bibfield  {journal} {\bibinfo
  {journal} {Physics of the earth and planetary interiors}\ }\textbf {\bibinfo
  {volume} {25}},\ \bibinfo {pages} {297--356} (\bibinfo {year}
  {1981})}\BibitemShut {NoStop}%
\bibitem [{\citenamefont {Gandhi}\ \emph {et~al.}(1996)\citenamefont {Gandhi},
  \citenamefont {Quigg}, \citenamefont {Reno},\ and\ \citenamefont
  {Sarcevic}}]{Gandhi:1995tf}%
  \BibitemOpen
  \bibfield  {author} {\bibinfo {author} {\bibfnamefont {Raj}\ \bibnamefont
  {Gandhi}}, \bibinfo {author} {\bibfnamefont {Chris}\ \bibnamefont {Quigg}},
  \bibinfo {author} {\bibfnamefont {Mary~Hall}\ \bibnamefont {Reno}}, \ and\
  \bibinfo {author} {\bibfnamefont {Ina}\ \bibnamefont {Sarcevic}},\ }\bibfield
   {title} {\enquote {\bibinfo {title} {{Ultrahigh-energy neutrino
  interactions}},}\ }\href {\doibase 10.1016/0927-6505(96)00008-4} {\bibfield
  {journal} {\bibinfo  {journal} {Astropart. Phys.}\ }\textbf {\bibinfo
  {volume} {5}},\ \bibinfo {pages} {81--110} (\bibinfo {year} {1996})},\
  \Eprint {http://arxiv.org/abs/hep-ph/9512364} {arXiv:hep-ph/9512364}
  \BibitemShut {NoStop}%
\bibitem [{\citenamefont {Abdul~Khalek}\ \emph {et~al.}(2022)\citenamefont
  {Abdul~Khalek}, \citenamefont {Gauld}, \citenamefont {Giani}, \citenamefont
  {Nocera}, \citenamefont {Rabemananjara},\ and\ \citenamefont
  {Rojo}}]{AbdulKhalek:2022fyi}%
  \BibitemOpen
  \bibfield  {author} {\bibinfo {author} {\bibfnamefont {Rabah}\ \bibnamefont
  {Abdul~Khalek}}, \bibinfo {author} {\bibfnamefont {Rhorry}\ \bibnamefont
  {Gauld}}, \bibinfo {author} {\bibfnamefont {Tommaso}\ \bibnamefont {Giani}},
  \bibinfo {author} {\bibfnamefont {Emanuele~R.}\ \bibnamefont {Nocera}},
  \bibinfo {author} {\bibfnamefont {Tanjona~R.}\ \bibnamefont {Rabemananjara}},
  \ and\ \bibinfo {author} {\bibfnamefont {Juan}\ \bibnamefont {Rojo}},\
  }\bibfield  {title} {\enquote {\bibinfo {title} {{nNNPDF3.0: evidence for a
  modified partonic structure in heavy nuclei}},}\ }\href {\doibase
  10.1140/epjc/s10052-022-10417-7} {\bibfield  {journal} {\bibinfo  {journal}
  {Eur. Phys. J. C}\ }\textbf {\bibinfo {volume} {82}},\ \bibinfo {pages} {507}
  (\bibinfo {year} {2022})},\ \Eprint {http://arxiv.org/abs/2201.12363}
  {arXiv:2201.12363 [hep-ph]} \BibitemShut {NoStop}%
\bibitem [{\citenamefont {Zhou}\ and\ \citenamefont
  {Beacom}(2020{\natexlab{a}})}]{Zhou:2019vxt}%
  \BibitemOpen
  \bibfield  {author} {\bibinfo {author} {\bibfnamefont {Bei}\ \bibnamefont
  {Zhou}}\ and\ \bibinfo {author} {\bibfnamefont {John~F.}\ \bibnamefont
  {Beacom}},\ }\bibfield  {title} {\enquote {\bibinfo {title}
  {{Neutrino-nucleus cross sections for W-boson and trident production}},}\
  }\href {\doibase 10.1103/PhysRevD.101.036011} {\bibfield  {journal} {\bibinfo
   {journal} {Phys. Rev. D}\ }\textbf {\bibinfo {volume} {101}},\ \bibinfo
  {pages} {036011} (\bibinfo {year} {2020}{\natexlab{a}})},\ \Eprint
  {http://arxiv.org/abs/1910.08090} {arXiv:1910.08090 [hep-ph]} \BibitemShut
  {NoStop}%
\bibitem [{\citenamefont {Zhou}\ and\ \citenamefont
  {Beacom}(2020{\natexlab{b}})}]{Zhou:2019frk}%
  \BibitemOpen
  \bibfield  {author} {\bibinfo {author} {\bibfnamefont {Bei}\ \bibnamefont
  {Zhou}}\ and\ \bibinfo {author} {\bibfnamefont {John~F.}\ \bibnamefont
  {Beacom}},\ }\bibfield  {title} {\enquote {\bibinfo {title} {{W-boson and
  trident production in TeV\textendash{}PeV neutrino observatories}},}\ }\href
  {\doibase 10.1103/PhysRevD.101.036010} {\bibfield  {journal} {\bibinfo
  {journal} {Phys. Rev. D}\ }\textbf {\bibinfo {volume} {101}},\ \bibinfo
  {pages} {036010} (\bibinfo {year} {2020}{\natexlab{b}})},\ \Eprint
  {http://arxiv.org/abs/1910.10720} {arXiv:1910.10720 [hep-ph]} \BibitemShut
  {NoStop}%
\bibitem [{\citenamefont {Garcia}\ \emph {et~al.}(2020)\citenamefont {Garcia},
  \citenamefont {Gauld}, \citenamefont {Heijboer},\ and\ \citenamefont
  {Rojo}}]{Garcia:2020jwr}%
  \BibitemOpen
  \bibfield  {author} {\bibinfo {author} {\bibfnamefont {Alfonso}\ \bibnamefont
  {Garcia}}, \bibinfo {author} {\bibfnamefont {Rhorry}\ \bibnamefont {Gauld}},
  \bibinfo {author} {\bibfnamefont {Aart}\ \bibnamefont {Heijboer}}, \ and\
  \bibinfo {author} {\bibfnamefont {Juan}\ \bibnamefont {Rojo}},\ }\bibfield
  {title} {\enquote {\bibinfo {title} {{Complete predictions for high-energy
  neutrino propagation in matter}},}\ }\href {\doibase
  10.1088/1475-7516/2020/09/025} {\bibfield  {journal} {\bibinfo  {journal}
  {JCAP}\ }\textbf {\bibinfo {volume} {09}},\ \bibinfo {pages} {025} (\bibinfo
  {year} {2020})},\ \Eprint {http://arxiv.org/abs/2004.04756} {arXiv:2004.04756
  [hep-ph]} \BibitemShut {NoStop}%
\bibitem [{\citenamefont {Zhou}\ and\ \citenamefont
  {Beacom}(2022)}]{Zhou:2021xuh}%
  \BibitemOpen
  \bibfield  {author} {\bibinfo {author} {\bibfnamefont {Bei}\ \bibnamefont
  {Zhou}}\ and\ \bibinfo {author} {\bibfnamefont {John~F.}\ \bibnamefont
  {Beacom}},\ }\bibfield  {title} {\enquote {\bibinfo {title} {{Dimuons in
  neutrino telescopes: New predictions and first search in IceCube}},}\ }\href
  {\doibase 10.1103/PhysRevD.105.093005} {\bibfield  {journal} {\bibinfo
  {journal} {Phys. Rev. D}\ }\textbf {\bibinfo {volume} {105}},\ \bibinfo
  {pages} {093005} (\bibinfo {year} {2022})},\ \Eprint
  {http://arxiv.org/abs/2110.02974} {arXiv:2110.02974 [hep-ph]} \BibitemShut
  {NoStop}%
\bibitem [{\citenamefont {Soto}\ \emph {et~al.}(2022)\citenamefont {Soto},
  \citenamefont {Zhelnin}, \citenamefont {Safa},\ and\ \citenamefont
  {Arg\"uelles}}]{Soto:2021vdc}%
  \BibitemOpen
  \bibfield  {author} {\bibinfo {author} {\bibfnamefont {Alfonso~Garcia}\
  \bibnamefont {Soto}}, \bibinfo {author} {\bibfnamefont {Pavel}\ \bibnamefont
  {Zhelnin}}, \bibinfo {author} {\bibfnamefont {Ibrahim}\ \bibnamefont {Safa}},
  \ and\ \bibinfo {author} {\bibfnamefont {Carlos~A.}\ \bibnamefont
  {Arg\"uelles}},\ }\bibfield  {title} {\enquote {\bibinfo {title} {{Tau
  Appearance from High-Energy Neutrino Interactions}},}\ }\href {\doibase
  10.1103/PhysRevLett.128.171101} {\bibfield  {journal} {\bibinfo  {journal}
  {Phys. Rev. Lett.}\ }\textbf {\bibinfo {volume} {128}},\ \bibinfo {pages}
  {171101} (\bibinfo {year} {2022})},\ \Eprint
  {http://arxiv.org/abs/2112.06937} {arXiv:2112.06937 [hep-ph]} \BibitemShut
  {NoStop}%
\bibitem [{\citenamefont {Ritz}\ and\ \citenamefont
  {Seckel}(1988)}]{Ritz:1987mh}%
  \BibitemOpen
  \bibfield  {author} {\bibinfo {author} {\bibfnamefont {S.}~\bibnamefont
  {Ritz}}\ and\ \bibinfo {author} {\bibfnamefont {D.}~\bibnamefont {Seckel}},\
  }\bibfield  {title} {\enquote {\bibinfo {title} {{Detailed Neutrino Spectra
  From Cold Dark Matter Annihilations in the Sun}},}\ }\href {\doibase
  10.1016/0550-3213(88)90660-8} {\bibfield  {journal} {\bibinfo  {journal}
  {Nucl. Phys. B}\ }\textbf {\bibinfo {volume} {304}},\ \bibinfo {pages}
  {877--908} (\bibinfo {year} {1988})}\BibitemShut {NoStop}%
\bibitem [{\citenamefont {Nicolaidis}\ and\ \citenamefont
  {Taramopoulos}(1996)}]{Nicolaidis:1996qu}%
  \BibitemOpen
  \bibfield  {author} {\bibinfo {author} {\bibfnamefont {A.}~\bibnamefont
  {Nicolaidis}}\ and\ \bibinfo {author} {\bibfnamefont {A.}~\bibnamefont
  {Taramopoulos}},\ }\bibfield  {title} {\enquote {\bibinfo {title} {{Shadowing
  of ultrahigh-energy neutrinos}},}\ }\href {\doibase
  10.1016/0370-2693(96)00948-3} {\bibfield  {journal} {\bibinfo  {journal}
  {Phys. Lett. B}\ }\textbf {\bibinfo {volume} {386}},\ \bibinfo {pages}
  {211--216} (\bibinfo {year} {1996})},\ \Eprint
  {http://arxiv.org/abs/hep-ph/9603382} {arXiv:hep-ph/9603382} \BibitemShut
  {NoStop}%
\bibitem [{\citenamefont {Halzen}\ and\ \citenamefont
  {Saltzberg}(1998)}]{Halzen:1998be}%
  \BibitemOpen
  \bibfield  {author} {\bibinfo {author} {\bibfnamefont {F.}~\bibnamefont
  {Halzen}}\ and\ \bibinfo {author} {\bibfnamefont {D.}~\bibnamefont
  {Saltzberg}},\ }\bibfield  {title} {\enquote {\bibinfo {title} {{Tau-neutrino
  appearance with a 1000 megaparsec baseline}},}\ }\href {\doibase
  10.1103/PhysRevLett.81.4305} {\bibfield  {journal} {\bibinfo  {journal}
  {Phys. Rev. Lett.}\ }\textbf {\bibinfo {volume} {81}},\ \bibinfo {pages}
  {4305--4308} (\bibinfo {year} {1998})},\ \Eprint
  {http://arxiv.org/abs/hep-ph/9804354} {arXiv:hep-ph/9804354} \BibitemShut
  {NoStop}%
\bibitem [{\citenamefont {Kwiecinski}\ \emph {et~al.}(1999)\citenamefont
  {Kwiecinski}, \citenamefont {Martin},\ and\ \citenamefont
  {Stasto}}]{Kwiecinski:1998yf}%
  \BibitemOpen
  \bibfield  {author} {\bibinfo {author} {\bibfnamefont {J.}~\bibnamefont
  {Kwiecinski}}, \bibinfo {author} {\bibfnamefont {Alan~D.}\ \bibnamefont
  {Martin}}, \ and\ \bibinfo {author} {\bibfnamefont {A.~M.}\ \bibnamefont
  {Stasto}},\ }\bibfield  {title} {\enquote {\bibinfo {title} {{Penetration of
  the earth by ultrahigh-energy neutrinos predicted by low x QCD}},}\ }\href
  {\doibase 10.1103/PhysRevD.59.093002} {\bibfield  {journal} {\bibinfo
  {journal} {Phys. Rev. D}\ }\textbf {\bibinfo {volume} {59}},\ \bibinfo
  {pages} {093002} (\bibinfo {year} {1999})},\ \Eprint
  {http://arxiv.org/abs/astro-ph/9812262} {arXiv:astro-ph/9812262} \BibitemShut
  {NoStop}%
\bibitem [{\citenamefont {Beacom}\ \emph {et~al.}(2002)\citenamefont {Beacom},
  \citenamefont {Crotty},\ and\ \citenamefont {Kolb}}]{Beacom:2001xn}%
  \BibitemOpen
  \bibfield  {author} {\bibinfo {author} {\bibfnamefont {John~F.}\ \bibnamefont
  {Beacom}}, \bibinfo {author} {\bibfnamefont {Patrick}\ \bibnamefont
  {Crotty}}, \ and\ \bibinfo {author} {\bibfnamefont {Edward~W.}\ \bibnamefont
  {Kolb}},\ }\bibfield  {title} {\enquote {\bibinfo {title} {{Enhanced Signal
  of Astrophysical Tau Neutrinos Propagating through Earth}},}\ }\href
  {\doibase 10.1103/PhysRevD.66.021302} {\bibfield  {journal} {\bibinfo
  {journal} {Phys. Rev. D}\ }\textbf {\bibinfo {volume} {66}},\ \bibinfo
  {pages} {021302(R)} (\bibinfo {year} {2002})},\ \Eprint
  {http://arxiv.org/abs/astro-ph/0111482} {arXiv:astro-ph/0111482} \BibitemShut
  {NoStop}%
\bibitem [{\citenamefont {Dutta}\ \emph {et~al.}(2002)\citenamefont {Dutta},
  \citenamefont {Reno},\ and\ \citenamefont {Sarcevic}}]{Dutta:2002zc}%
  \BibitemOpen
  \bibfield  {author} {\bibinfo {author} {\bibfnamefont {Sharada~Iyer}\
  \bibnamefont {Dutta}}, \bibinfo {author} {\bibfnamefont {Mary~Hall}\
  \bibnamefont {Reno}}, \ and\ \bibinfo {author} {\bibfnamefont {Ina}\
  \bibnamefont {Sarcevic}},\ }\bibfield  {title} {\enquote {\bibinfo {title}
  {{Secondary neutrinos from tau neutrino interactions in earth}},}\ }\href
  {\doibase 10.1103/PhysRevD.66.077302} {\bibfield  {journal} {\bibinfo
  {journal} {Phys. Rev. D}\ }\textbf {\bibinfo {volume} {66}},\ \bibinfo
  {pages} {077302} (\bibinfo {year} {2002})},\ \Eprint
  {http://arxiv.org/abs/hep-ph/0207344} {arXiv:hep-ph/0207344} \BibitemShut
  {NoStop}%
\bibitem [{\citenamefont {Arg\"uelles}\ \emph {et~al.}(2022)\citenamefont
  {Arg\"uelles}, \citenamefont {Halzen}, \citenamefont {Kheirandish},\ and\
  \citenamefont {Safa}}]{Arguelles:2022aum}%
  \BibitemOpen
  \bibfield  {author} {\bibinfo {author} {\bibfnamefont {Carlos~A.}\
  \bibnamefont {Arg\"uelles}}, \bibinfo {author} {\bibfnamefont {Francis}\
  \bibnamefont {Halzen}}, \bibinfo {author} {\bibfnamefont {Ali}\ \bibnamefont
  {Kheirandish}}, \ and\ \bibinfo {author} {\bibfnamefont {Ibrahim}\
  \bibnamefont {Safa}},\ }\bibfield  {title} {\enquote {\bibinfo {title} {{PeV
  Tau Neutrinos to Unveil Ultra-High-Energy Sources}},}\ }\href@noop {} {\
  (\bibinfo {year} {2022})},\ \Eprint {http://arxiv.org/abs/2203.13827}
  {arXiv:2203.13827 [astro-ph.HE]} \BibitemShut {NoStop}%
\bibitem [{\citenamefont {Allison}\ \emph {et~al.}(2022)\citenamefont {Allison}
  \emph {et~al.}}]{ARA:2022rwq}%
  \BibitemOpen
  \bibfield  {author} {\bibinfo {author} {\bibfnamefont {P.}~\bibnamefont
  {Allison}} \emph {et~al.} (\bibinfo {collaboration} {ARA}),\ }\bibfield
  {title} {\enquote {\bibinfo {title} {{Low-threshold ultrahigh-energy neutrino
  search with the Askaryan Radio Array}},}\ }\href {\doibase
  10.1103/PhysRevD.105.122006} {\bibfield  {journal} {\bibinfo  {journal}
  {Phys. Rev. D}\ }\textbf {\bibinfo {volume} {105}},\ \bibinfo {pages}
  {122006} (\bibinfo {year} {2022})},\ \Eprint
  {http://arxiv.org/abs/2202.07080} {arXiv:2202.07080 [astro-ph.HE]}
  \BibitemShut {NoStop}%
\bibitem [{\citenamefont {Jain}\ \emph {et~al.}(2000)\citenamefont {Jain},
  \citenamefont {McKay}, \citenamefont {Panda},\ and\ \citenamefont
  {Ralston}}]{Jain:2000pu}%
  \BibitemOpen
  \bibfield  {author} {\bibinfo {author} {\bibfnamefont {P.}~\bibnamefont
  {Jain}}, \bibinfo {author} {\bibfnamefont {Douglas~W.}\ \bibnamefont
  {McKay}}, \bibinfo {author} {\bibfnamefont {S.}~\bibnamefont {Panda}}, \ and\
  \bibinfo {author} {\bibfnamefont {John~P.}\ \bibnamefont {Ralston}},\
  }\bibfield  {title} {\enquote {\bibinfo {title} {{Extra dimensions and strong
  neutrino nucleon interactions above $10^{19}$ eV: Breaking the GZK
  barrier}},}\ }\href {\doibase 10.1016/S0370-2693(00)00647-X} {\bibfield
  {journal} {\bibinfo  {journal} {Phys. Lett. B}\ }\textbf {\bibinfo {volume}
  {484}},\ \bibinfo {pages} {267--274} (\bibinfo {year} {2000})},\ \Eprint
  {http://arxiv.org/abs/hep-ph/0001031} {arXiv:hep-ph/0001031} \BibitemShut
  {NoStop}%
\bibitem [{\citenamefont {Be\v{c}irevi\'c}\ \emph {et~al.}(2018)\citenamefont
  {Be\v{c}irevi\'c}, \citenamefont {Panes}, \citenamefont {Sumensari},\ and\
  \citenamefont {Zukanovich~Funchal}}]{Becirevic:2018uab}%
  \BibitemOpen
  \bibfield  {author} {\bibinfo {author} {\bibfnamefont {Damir}\ \bibnamefont
  {Be\v{c}irevi\'c}}, \bibinfo {author} {\bibfnamefont {Boris}\ \bibnamefont
  {Panes}}, \bibinfo {author} {\bibfnamefont {Olcyr}\ \bibnamefont
  {Sumensari}}, \ and\ \bibinfo {author} {\bibfnamefont {Renata}\ \bibnamefont
  {Zukanovich~Funchal}},\ }\bibfield  {title} {\enquote {\bibinfo {title}
  {{Seeking leptoquarks in IceCube}},}\ }\href {\doibase
  10.1007/JHEP06(2018)032} {\bibfield  {journal} {\bibinfo  {journal} {JHEP}\
  }\textbf {\bibinfo {volume} {06}},\ \bibinfo {pages} {032} (\bibinfo {year}
  {2018})},\ \Eprint {http://arxiv.org/abs/1803.10112} {arXiv:1803.10112
  [hep-ph]} \BibitemShut {NoStop}%
\bibitem [{\citenamefont {Arg\"uelles}\ \emph {et~al.}(2015)\citenamefont
  {Arg\"uelles}, \citenamefont {Halzen}, \citenamefont {Wille}, \citenamefont
  {Kroll},\ and\ \citenamefont {Reno}}]{Arguelles:2015wba}%
  \BibitemOpen
  \bibfield  {author} {\bibinfo {author} {\bibfnamefont {Carlos~A.}\
  \bibnamefont {Arg\"uelles}}, \bibinfo {author} {\bibfnamefont {Francis}\
  \bibnamefont {Halzen}}, \bibinfo {author} {\bibfnamefont {Logan}\
  \bibnamefont {Wille}}, \bibinfo {author} {\bibfnamefont {Mike}\ \bibnamefont
  {Kroll}}, \ and\ \bibinfo {author} {\bibfnamefont {Mary~Hall}\ \bibnamefont
  {Reno}},\ }\bibfield  {title} {\enquote {\bibinfo {title} {{High-energy
  behavior of photon, neutrino, and proton cross sections}},}\ }\href {\doibase
  10.1103/PhysRevD.92.074040} {\bibfield  {journal} {\bibinfo  {journal} {Phys.
  Rev. D}\ }\textbf {\bibinfo {volume} {92}},\ \bibinfo {pages} {074040}
  (\bibinfo {year} {2015})},\ \Eprint {http://arxiv.org/abs/1504.06639}
  {arXiv:1504.06639 [hep-ph]} \BibitemShut {NoStop}%
\bibitem [{\citenamefont {van Santen}\ \emph {et~al.}(2022)\citenamefont {van
  Santen}, \citenamefont {Clark}, \citenamefont {Halliday}, \citenamefont
  {Hallmann},\ and\ \citenamefont {Nelles}}]{vanSanten:2022wss}%
  \BibitemOpen
  \bibfield  {author} {\bibinfo {author} {\bibfnamefont {Jakob}\ \bibnamefont
  {van Santen}}, \bibinfo {author} {\bibfnamefont {Brian~A.}\ \bibnamefont
  {Clark}}, \bibinfo {author} {\bibfnamefont {Rob}\ \bibnamefont {Halliday}},
  \bibinfo {author} {\bibfnamefont {Steffen}\ \bibnamefont {Hallmann}}, \ and\
  \bibinfo {author} {\bibfnamefont {Anna}\ \bibnamefont {Nelles}},\ }\bibfield
  {title} {\enquote {\bibinfo {title} {{toise: a framework to describe the
  performance of high-energy neutrino detectors}},}\ }\href@noop {} {\
  (\bibinfo {year} {2022})},\ \Eprint {http://arxiv.org/abs/2202.11120}
  {arXiv:2202.11120 [astro-ph.IM]} \BibitemShut {NoStop}%
\bibitem [{\citenamefont {Ackermann}\ \emph {et~al.}(2022)\citenamefont
  {Ackermann} \emph {et~al.}}]{Ackermann:2022rqc}%
  \BibitemOpen
  \bibfield  {author} {\bibinfo {author} {\bibfnamefont {Markus}\ \bibnamefont
  {Ackermann}} \emph {et~al.},\ }\bibfield  {title} {\enquote {\bibinfo {title}
  {{High-Energy and Ultra-High-Energy Neutrinos}},}\ }\href@noop {} {\
  (\bibinfo {year} {2022})},\ \Eprint {http://arxiv.org/abs/2203.08096}
  {arXiv:2203.08096 [hep-ph]} \BibitemShut {NoStop}%
\bibitem [{\citenamefont {Kravchenko}\ \emph {et~al.}(2006)\citenamefont
  {Kravchenko} \emph {et~al.}}]{Kravchenko:2006qc}%
  \BibitemOpen
  \bibfield  {author} {\bibinfo {author} {\bibfnamefont {I.}~\bibnamefont
  {Kravchenko}} \emph {et~al.},\ }\bibfield  {title} {\enquote {\bibinfo
  {title} {{Rice limits on the diffuse ultrahigh energy neutrino flux}},}\
  }\href {\doibase 10.1103/PhysRevD.73.082002} {\bibfield  {journal} {\bibinfo
  {journal} {Phys. Rev. D}\ }\textbf {\bibinfo {volume} {73}},\ \bibinfo
  {pages} {082002} (\bibinfo {year} {2006})},\ \Eprint
  {http://arxiv.org/abs/astro-ph/0601148} {arXiv:astro-ph/0601148} \BibitemShut
  {NoStop}%
\bibitem [{\citenamefont {de~Vries}\ \emph {et~al.}(2016)\citenamefont
  {de~Vries}, \citenamefont {Buitink}, \citenamefont {van Eijndhoven},
  \citenamefont {Meures}, \citenamefont {\'O~Murchadha},\ and\ \citenamefont
  {Scholten}}]{deVries:2015oda}%
  \BibitemOpen
  \bibfield  {author} {\bibinfo {author} {\bibfnamefont {Krijn~D.}\
  \bibnamefont {de~Vries}}, \bibinfo {author} {\bibfnamefont {Stijn}\
  \bibnamefont {Buitink}}, \bibinfo {author} {\bibfnamefont {Nick}\
  \bibnamefont {van Eijndhoven}}, \bibinfo {author} {\bibfnamefont {Thomas}\
  \bibnamefont {Meures}}, \bibinfo {author} {\bibfnamefont {Aongus}\
  \bibnamefont {\'O~Murchadha}}, \ and\ \bibinfo {author} {\bibfnamefont
  {Olaf}\ \bibnamefont {Scholten}},\ }\bibfield  {title} {\enquote {\bibinfo
  {title} {{The cosmic-ray air-shower signal in Askaryan radio detectors}},}\
  }\href {\doibase 10.1016/j.astropartphys.2015.10.003} {\bibfield  {journal}
  {\bibinfo  {journal} {Astropart. Phys.}\ }\textbf {\bibinfo {volume} {74}},\
  \bibinfo {pages} {96--104} (\bibinfo {year} {2016})},\ \Eprint
  {http://arxiv.org/abs/1503.02808} {arXiv:1503.02808 [astro-ph.HE]}
  \BibitemShut {NoStop}%
\bibitem [{\citenamefont {Palomares-Ruiz}\ \emph {et~al.}(2006)\citenamefont
  {Palomares-Ruiz}, \citenamefont {Irimia},\ and\ \citenamefont
  {Weiler}}]{Palomares-Ruiz:2005npx}%
  \BibitemOpen
  \bibfield  {author} {\bibinfo {author} {\bibfnamefont {Sergio}\ \bibnamefont
  {Palomares-Ruiz}}, \bibinfo {author} {\bibfnamefont {Andrei}\ \bibnamefont
  {Irimia}}, \ and\ \bibinfo {author} {\bibfnamefont {Thomas~J.}\ \bibnamefont
  {Weiler}},\ }\bibfield  {title} {\enquote {\bibinfo {title} {{Acceptances for
  space-based and ground-based fluorescence detectors, and inference of the
  neutrino-nucleon cross-section above 10**19-ev}},}\ }\href {\doibase
  10.1103/PhysRevD.73.083003} {\bibfield  {journal} {\bibinfo  {journal} {Phys.
  Rev. D}\ }\textbf {\bibinfo {volume} {73}},\ \bibinfo {pages} {083003}
  (\bibinfo {year} {2006})},\ \Eprint {http://arxiv.org/abs/astro-ph/0512231}
  {arXiv:astro-ph/0512231} \BibitemShut {NoStop}%
\bibitem [{\citenamefont {Pauli}(1930)}]{Pauli:1930pc}%
  \BibitemOpen
  \bibfield  {author} {\bibinfo {author} {\bibfnamefont {W.}~\bibnamefont
  {Pauli}},\ }\bibfield  {title} {\enquote {\bibinfo {title} {{Dear radioactive
  ladies and gentlemen}},}\ }\href@noop {} {\  (\bibinfo {year} {1930})},\
  \bibinfo {note} {{reproduced in Phys. Today \textbf{31N9}, 27
  (1978)}}\BibitemShut {NoStop}%
\bibitem [{\citenamefont {Bethe}\ and\ \citenamefont
  {Peierls}(1934)}]{Bethe:1934qn}%
  \BibitemOpen
  \bibfield  {author} {\bibinfo {author} {\bibfnamefont {H.}~\bibnamefont
  {Bethe}}\ and\ \bibinfo {author} {\bibfnamefont {R.}~\bibnamefont
  {Peierls}},\ }\bibfield  {title} {\enquote {\bibinfo {title} {{The
  'neutrino'}},}\ }\href {\doibase 10.1038/133532a0} {\bibfield  {journal}
  {\bibinfo  {journal} {Nature}\ }\textbf {\bibinfo {volume} {133}},\ \bibinfo
  {pages} {532} (\bibinfo {year} {1934})}\BibitemShut {NoStop}%
\bibitem [{\citenamefont {Besson}\ \emph {et~al.}(2021)\citenamefont {Besson},
  \citenamefont {Kravchenko},\ and\ \citenamefont {Nivedita}}]{Besson:2021wmj}%
  \BibitemOpen
  \bibfield  {author} {\bibinfo {author} {\bibfnamefont {Dave~Z.}\ \bibnamefont
  {Besson}}, \bibinfo {author} {\bibfnamefont {Ilya}\ \bibnamefont
  {Kravchenko}}, \ and\ \bibinfo {author} {\bibfnamefont {Krishna}\
  \bibnamefont {Nivedita}},\ }\bibfield  {title} {\enquote {\bibinfo {title}
  {{Angular Dependence of Vertically Propagating Radio-Frequency Signals in
  South Polar Ice}},}\ }\href@noop {} {\  (\bibinfo {year} {2021})},\ \Eprint
  {http://arxiv.org/abs/2110.13353} {arXiv:2110.13353 [astro-ph.IM]}
  \BibitemShut {NoStop}%
\bibitem [{\citenamefont {Allison}\ \emph
  {et~al.}(2020{\natexlab{b}})\citenamefont {Allison} \emph
  {et~al.}}]{Allison:2019rgg}%
  \BibitemOpen
  \bibfield  {author} {\bibinfo {author} {\bibfnamefont {P.}~\bibnamefont
  {Allison}} \emph {et~al.},\ }\bibfield  {title} {\enquote {\bibinfo {title}
  {{Long-baseline horizontal radio-frequency transmission through polar
  ice}},}\ }\href {\doibase 10.1088/1475-7516/2020/12/009} {\bibfield
  {journal} {\bibinfo  {journal} {JCAP}\ }\textbf {\bibinfo {volume} {12}},\
  \bibinfo {pages} {009} (\bibinfo {year} {2020}{\natexlab{b}})},\ \Eprint
  {http://arxiv.org/abs/1908.10689} {arXiv:1908.10689 [astro-ph.IM]}
  \BibitemShut {NoStop}%
\bibitem [{\citenamefont {Jordan}\ \emph {et~al.}(2020)\citenamefont {Jordan},
  \citenamefont {Besson}, \citenamefont {Kravchenko}, \citenamefont {Latif},
  \citenamefont {Madison}, \citenamefont {Novikov},\ and\ \citenamefont
  {Shultz}}]{Jordan:2019bqu}%
  \BibitemOpen
  \bibfield  {author} {\bibinfo {author} {\bibfnamefont {T.~M.}\ \bibnamefont
  {Jordan}}, \bibinfo {author} {\bibfnamefont {D.~Z.}\ \bibnamefont {Besson}},
  \bibinfo {author} {\bibfnamefont {I.}~\bibnamefont {Kravchenko}}, \bibinfo
  {author} {\bibfnamefont {U.}~\bibnamefont {Latif}}, \bibinfo {author}
  {\bibfnamefont {B.}~\bibnamefont {Madison}}, \bibinfo {author} {\bibfnamefont
  {A.}~\bibnamefont {Novikov}}, \ and\ \bibinfo {author} {\bibfnamefont
  {A.}~\bibnamefont {Shultz}},\ }\bibfield  {title} {\enquote {\bibinfo {title}
  {{Modelling ice birefringence and oblique radio wave propagation for neutrino
  detection at the South Pole}},}\ }\href {\doibase 10.1017/aog.2020.18}
  {\bibfield  {journal} {\bibinfo  {journal} {Annals Glaciol.}\ }\textbf
  {\bibinfo {volume} {61}},\ \bibinfo {pages} {84} (\bibinfo {year} {2020})},\
  \Eprint {http://arxiv.org/abs/1910.01471} {arXiv:1910.01471 [astro-ph.IM]}
  \BibitemShut {NoStop}%
\bibitem [{\citenamefont {Connolly}(2022)}]{Connolly:2021cum}%
  \BibitemOpen
  \bibfield  {author} {\bibinfo {author} {\bibfnamefont {Amy}\ \bibnamefont
  {Connolly}},\ }\bibfield  {title} {\enquote {\bibinfo {title} {{Impact of
  biaxial birefringence in polar ice at radio frequencies on signal
  polarizations in ultrahigh energy neutrino detection}},}\ }\href {\doibase
  10.1103/PhysRevD.105.123012} {\bibfield  {journal} {\bibinfo  {journal}
  {Phys. Rev. D}\ }\textbf {\bibinfo {volume} {105}},\ \bibinfo {pages}
  {123012} (\bibinfo {year} {2022})},\ \Eprint
  {http://arxiv.org/abs/2110.09015} {arXiv:2110.09015 [astro-ph.IM]}
  \BibitemShut {NoStop}%
\bibitem [{\citenamefont {Aguilar}\ \emph {et~al.}(2022)\citenamefont {Aguilar}
  \emph {et~al.}}]{Aguilar:2022kgi}%
  \BibitemOpen
  \bibfield  {author} {\bibinfo {author} {\bibfnamefont {J.~A.}\ \bibnamefont
  {Aguilar}} \emph {et~al.},\ }\bibfield  {title} {\enquote {\bibinfo {title}
  {{In situ, broadband measurement of the radio frequency attenuation length at
  Summit Station, Greenland}},}\ }\href@noop {} {\  (\bibinfo {year} {2022})},\
  \Eprint {http://arxiv.org/abs/2201.07846} {arXiv:2201.07846 [astro-ph.IM]}
  \BibitemShut {NoStop}%
\bibitem [{\citenamefont {Prohira}\ \emph
  {et~al.}(2021{\natexlab{b}})\citenamefont {Prohira} \emph
  {et~al.}}]{RadarEchoTelescope:2020nhe}%
  \BibitemOpen
  \bibfield  {author} {\bibinfo {author} {\bibfnamefont {S.}~\bibnamefont
  {Prohira}} \emph {et~al.} (\bibinfo {collaboration} {Radar Echo Telescope}),\
  }\bibfield  {title} {\enquote {\bibinfo {title} {{Modeling in-ice radio
  propagation with parabolic equation methods}},}\ }\href {\doibase
  10.1103/PhysRevD.103.103007} {\bibfield  {journal} {\bibinfo  {journal}
  {Phys. Rev. D}\ }\textbf {\bibinfo {volume} {103}},\ \bibinfo {pages}
  {103007} (\bibinfo {year} {2021}{\natexlab{b}})},\ \Eprint
  {http://arxiv.org/abs/2011.05997} {arXiv:2011.05997 [astro-ph.IM]}
  \BibitemShut {NoStop}%
\bibitem [{\citenamefont {Anker}\ \emph {et~al.}(2020)\citenamefont {Anker}
  \emph {et~al.}}]{ARIANNA:2020zrg}%
  \BibitemOpen
  \bibfield  {author} {\bibinfo {author} {\bibfnamefont {A.}~\bibnamefont
  {Anker}} \emph {et~al.} (\bibinfo {collaboration} {ARIANNA}),\ }\bibfield
  {title} {\enquote {\bibinfo {title} {{Probing the angular and polarization
  reconstruction of the ARIANNA detector at the South Pole}},}\ }\href
  {\doibase 10.1088/1748-0221/15/09/P09039} {\bibfield  {journal} {\bibinfo
  {journal} {JINST}\ }\textbf {\bibinfo {volume} {15}},\ \bibinfo {pages}
  {P09039} (\bibinfo {year} {2020})},\ \Eprint
  {http://arxiv.org/abs/2006.03027} {arXiv:2006.03027 [astro-ph.IM]}
  \BibitemShut {NoStop}%
\bibitem [{\citenamefont {Fermi}(1949)}]{Fermi:1949ee}%
  \BibitemOpen
  \bibfield  {author} {\bibinfo {author} {\bibfnamefont {Enrico}\ \bibnamefont
  {Fermi}},\ }\bibfield  {title} {\enquote {\bibinfo {title} {{On the Origin of
  the Cosmic Radiation}},}\ }\href {\doibase 10.1103/PhysRev.75.1169}
  {\bibfield  {journal} {\bibinfo  {journal} {Phys. Rev.}\ }\textbf {\bibinfo
  {volume} {75}},\ \bibinfo {pages} {1169--1174} (\bibinfo {year}
  {1949})}\BibitemShut {NoStop}%
\bibitem [{\citenamefont {Gaisser}\ \emph {et~al.}(2016)\citenamefont
  {Gaisser}, \citenamefont {Engel},\ and\ \citenamefont
  {Resconi}}]{Gaisser:2016uoy}%
  \BibitemOpen
  \bibfield  {author} {\bibinfo {author} {\bibfnamefont {Thomas~K.}\
  \bibnamefont {Gaisser}}, \bibinfo {author} {\bibfnamefont {Ralph}\
  \bibnamefont {Engel}}, \ and\ \bibinfo {author} {\bibfnamefont {Elisa}\
  \bibnamefont {Resconi}},\ }\href@noop {} {\emph {\bibinfo {title} {{Cosmic
  Rays and Particle Physics}: {2nd Edition}}}}\ (\bibinfo  {publisher}
  {Cambridge University Press},\ \bibinfo {year} {2016})\BibitemShut {NoStop}%
\bibitem [{\citenamefont {Mégnin}\ and\ \citenamefont
  {Romanowicz}(2000)}]{Megnin:2000}%
  \BibitemOpen
  \bibfield  {author} {\bibinfo {author} {\bibfnamefont {Charles}\ \bibnamefont
  {Mégnin}}\ and\ \bibinfo {author} {\bibfnamefont {Barbara}\ \bibnamefont
  {Romanowicz}},\ }\bibfield  {title} {\enquote {\bibinfo {title} {{The
  three‐dimensional shear velocity structure of the mantle from the inversion
  of body, surface and higher‐mode waveforms}},}\ }\href {\doibase
  10.1046/j.1365-246X.2000.00298.x} {\bibfield  {journal} {\bibinfo  {journal}
  {Geophysical Journal International}\ }\textbf {\bibinfo {volume} {143}},\
  \bibinfo {pages} {709--728} (\bibinfo {year} {2000})}\BibitemShut {NoStop}%
\bibitem [{\citenamefont {Shen}\ and\ \citenamefont
  {Ritzwoller}(2016)}]{Shen:2016kxw}%
  \BibitemOpen
  \bibfield  {author} {\bibinfo {author} {\bibfnamefont {Weisen}\ \bibnamefont
  {Shen}}\ and\ \bibinfo {author} {\bibfnamefont {Michael~H.}\ \bibnamefont
  {Ritzwoller}},\ }\bibfield  {title} {\enquote {\bibinfo {title} {{Crustal and
  uppermost mantle structure beneath the United States}},}\ }\href {\doibase
  10.1002/2016jb012887} {\bibfield  {journal} {\bibinfo  {journal} {Journal of
  Geophysical Research: Solid Earth}\ }\textbf {\bibinfo {volume} {121}},\
  \bibinfo {pages} {4306--4342} (\bibinfo {year} {2016})}\BibitemShut {NoStop}%
\bibitem [{\citenamefont {Tao}\ \emph {et~al.}(2018)\citenamefont {Tao},
  \citenamefont {Grand},\ and\ \citenamefont {Niu}}]{Kai:2018}%
  \BibitemOpen
  \bibfield  {author} {\bibinfo {author} {\bibfnamefont {Kai}\ \bibnamefont
  {Tao}}, \bibinfo {author} {\bibfnamefont {Stephen~P.}\ \bibnamefont {Grand}},
  \ and\ \bibinfo {author} {\bibfnamefont {Fenglin}\ \bibnamefont {Niu}},\
  }\bibfield  {title} {\enquote {\bibinfo {title} {Seismic structure of the
  upper mantle beneath eastern asia from full waveform seismic tomography},}\
  }\href {\doibase https://doi.org/10.1029/2018GC007460} {\bibfield  {journal}
  {\bibinfo  {journal} {Geochemistry, Geophysics, Geosystems}\ }\textbf
  {\bibinfo {volume} {19}},\ \bibinfo {pages} {2732--2763} (\bibinfo {year}
  {2018})}\BibitemShut {NoStop}%
\bibitem [{\citenamefont {{Laske}}\ \emph {et~al.}(2013)\citenamefont
  {{Laske}}, \citenamefont {{Masters}}, \citenamefont {{Ma}},\ and\
  \citenamefont {{Pasyanos}}}]{Laske:2013}%
  \BibitemOpen
  \bibfield  {author} {\bibinfo {author} {\bibfnamefont {Gabi}\ \bibnamefont
  {{Laske}}}, \bibinfo {author} {\bibfnamefont {Guy}\ \bibnamefont
  {{Masters}}}, \bibinfo {author} {\bibfnamefont {Zhitu}\ \bibnamefont {{Ma}}},
  \ and\ \bibinfo {author} {\bibfnamefont {Mike}\ \bibnamefont {{Pasyanos}}},\
  }\bibfield  {title} {\enquote {\bibinfo {title} {{Update on CRUST1.0 - A
  1-degree Global Model of Earth's Crust}},}\ }in\ \href@noop {} {\emph
  {\bibinfo {booktitle} {EGU General Assembly Conference Abstracts}}},\
  \bibinfo {series and number} {EGU General Assembly Conference Abstracts}\
  (\bibinfo {year} {2013})\ pp.\ \bibinfo {pages} {EGU2013--2658}\BibitemShut
  {NoStop}%
\bibitem [{\citenamefont {Fretwell}\ \emph {et~al.}(2013)\citenamefont
  {Fretwell} \emph {et~al.}}]{Fretwell:2013}%
  \BibitemOpen
  \bibfield  {author} {\bibinfo {author} {\bibfnamefont {P.}~\bibnamefont
  {Fretwell}} \emph {et~al.},\ }\bibfield  {title} {\enquote {\bibinfo {title}
  {{Bedmap2: improved ice bed, surface and thickness datasets for
  Antarctica}},}\ }\href {\doibase 10.5194/tc-7-375-2013} {\bibfield  {journal}
  {\bibinfo  {journal} {The Cryosphere}\ }\textbf {\bibinfo {volume} {7}},\
  \bibinfo {pages} {375--393} (\bibinfo {year} {2013})}\BibitemShut {NoStop}%
\bibitem [{\citenamefont {Shen}\ \emph {et~al.}(2018)\citenamefont {Shen} \emph
  {et~al.}}]{Weisen:2018}%
  \BibitemOpen
  \bibfield  {author} {\bibinfo {author} {\bibfnamefont {Weisen}\ \bibnamefont
  {Shen}} \emph {et~al.},\ }\bibfield  {title} {\enquote {\bibinfo {title}
  {{The Crust and Upper Mantle Structure of Central and West Antarctica From
  Bayesian Inversion of Rayleigh Wave and Receiver Functions}},}\ }\href
  {\doibase https://doi.org/10.1029/2017JB015346} {\bibfield  {journal}
  {\bibinfo  {journal} {Journal of Geophysical Research: Solid Earth}\ }\textbf
  {\bibinfo {volume} {123}},\ \bibinfo {pages} {7824--7849} (\bibinfo {year}
  {2018})}\BibitemShut {NoStop}%
\bibitem [{\citenamefont {Bakhti}\ and\ \citenamefont
  {Smirnov}(2020)}]{Bakhti:2020tcj}%
  \BibitemOpen
  \bibfield  {author} {\bibinfo {author} {\bibfnamefont {Pouya}\ \bibnamefont
  {Bakhti}}\ and\ \bibinfo {author} {\bibfnamefont {Alexei~Yu.}\ \bibnamefont
  {Smirnov}},\ }\bibfield  {title} {\enquote {\bibinfo {title} {{Oscillation
  tomography of the Earth with solar neutrinos and future experiments}},}\
  }\href {\doibase 10.1103/PhysRevD.101.123031} {\bibfield  {journal} {\bibinfo
   {journal} {Phys. Rev. D}\ }\textbf {\bibinfo {volume} {101}},\ \bibinfo
  {pages} {123031} (\bibinfo {year} {2020})},\ \Eprint
  {http://arxiv.org/abs/2001.08030} {arXiv:2001.08030 [hep-ph]} \BibitemShut
  {NoStop}%
\bibitem [{\citenamefont {Safa}\ \emph {et~al.}(2020)\citenamefont {Safa},
  \citenamefont {Pizzuto}, \citenamefont {Arg\"uelles}, \citenamefont {Halzen},
  \citenamefont {Hussain}, \citenamefont {Kheirandish},\ and\ \citenamefont
  {Vandenbroucke}}]{Safa:2019ege}%
  \BibitemOpen
  \bibfield  {author} {\bibinfo {author} {\bibfnamefont {Ibrahim}\ \bibnamefont
  {Safa}}, \bibinfo {author} {\bibfnamefont {Alex}\ \bibnamefont {Pizzuto}},
  \bibinfo {author} {\bibfnamefont {Carlos~A.}\ \bibnamefont {Arg\"uelles}},
  \bibinfo {author} {\bibfnamefont {Francis}\ \bibnamefont {Halzen}}, \bibinfo
  {author} {\bibfnamefont {Raamis}\ \bibnamefont {Hussain}}, \bibinfo {author}
  {\bibfnamefont {Ali}\ \bibnamefont {Kheirandish}}, \ and\ \bibinfo {author}
  {\bibfnamefont {Justin}\ \bibnamefont {Vandenbroucke}},\ }\bibfield  {title}
  {\enquote {\bibinfo {title} {{Observing EeV neutrinos through Earth: GZK and
  the anomalous ANITA events}},}\ }\href {\doibase
  10.1088/1475-7516/2020/01/012} {\bibfield  {journal} {\bibinfo  {journal}
  {JCAP}\ }\textbf {\bibinfo {volume} {01}},\ \bibinfo {pages} {012} (\bibinfo
  {year} {2020})},\ \Eprint {http://arxiv.org/abs/1909.10487} {arXiv:1909.10487
  [hep-ph]} \BibitemShut {NoStop}%
\bibitem [{\citenamefont {Safa}\ \emph {et~al.}(2022)\citenamefont {Safa},
  \citenamefont {Lazar}, \citenamefont {Pizzuto}, \citenamefont {Vasquez},
  \citenamefont {Arg\"uelles},\ and\ \citenamefont
  {Vandenbroucke}}]{Safa:2021ghs}%
  \BibitemOpen
  \bibfield  {author} {\bibinfo {author} {\bibfnamefont {Ibrahim}\ \bibnamefont
  {Safa}}, \bibinfo {author} {\bibfnamefont {Jeffrey}\ \bibnamefont {Lazar}},
  \bibinfo {author} {\bibfnamefont {Alex}\ \bibnamefont {Pizzuto}}, \bibinfo
  {author} {\bibfnamefont {Oswaldo}\ \bibnamefont {Vasquez}}, \bibinfo {author}
  {\bibfnamefont {Carlos~A.}\ \bibnamefont {Arg\"uelles}}, \ and\ \bibinfo
  {author} {\bibfnamefont {Justin}\ \bibnamefont {Vandenbroucke}},\ }\bibfield
  {title} {\enquote {\bibinfo {title} {{TauRunner: A public Python program to
  propagate neutral and charged leptons}},}\ }\href {\doibase
  10.1016/j.cpc.2022.108422} {\bibfield  {journal} {\bibinfo  {journal}
  {Comput. Phys. Commun.}\ }\textbf {\bibinfo {volume} {278}},\ \bibinfo
  {pages} {108422} (\bibinfo {year} {2022})},\ \Eprint
  {http://arxiv.org/abs/2110.14662} {arXiv:2110.14662 [hep-ph]} \BibitemShut
  {NoStop}%
\end{thebibliography}%


\clearpage
\onecolumngrid
\appendix
\counterwithin{figure}{section}

\centerline{\Large {\bf Appendices}}
\bigskip

Here we provide more details and extended results, which may help support further developments. First, we describe how we carry out our sensitivity forecasts in \cref{appendix:fit}. We then extend our main results: in \cref{appendix:energies} to different neutrino energies and in \cref{appendix:flux} to different astrophysical fluxes. Finally, we show that subdominant effects are negligible: the specific Earth density profile in \cref{appendix:ice}, and neutrino regeneration in \cref{appendix:tau}.


\section{Details of the analysis}
\label{appendix:fit}

Here we describe in detail our procedure to obtain the cross-section sensitivity given in the main text.  In general, we denote as ${\mu(E_\mathrm{rec}, \theta_\mathrm{rec}) \, \mathrm{d}E_\mathrm{rec} \mathrm{d}\theta_\mathrm{rec}}$ the expected number of events with reconstructed energy between $E_\mathrm{rec}$ and $E_\mathrm{rec} + \mathrm{d}E_\mathrm{rec}$, and reconstructed angle between $\theta_\mathrm{rec}$ and $\theta_\mathrm{rec} + \mathrm{d}\theta_\mathrm{rec}$. This is given by (with definitions following)
\begin{equation}
    \mu(E_\mathrm{rec}, \theta_\mathrm{rec}) = N_\mathrm{nuc} \Delta t \int \mathrm{d}E_\nu \int \mathrm{d}\Omega \, \frac{\mathrm{d}\phi}{\mathrm{d}E_\nu \mathrm{d}\Omega \mathrm{d}t \mathrm{d}A} \sigma(E_\nu)  e^{-nL(\theta) \sigma(E_\nu)} \, R(E_\mathrm{rec}, E_\nu)  R(\theta_\mathrm{rec}, \theta)   \varepsilon(E_\nu) \, ,
    \label{eq:expected}
\end{equation}
\begin{itemize}

    \item $N_\mathrm{nuc}$ is the total number of nucleons in the sensitive volume.
    It is related to the effective volume $V_\mathrm{eff}$ by ${V_\mathrm{eff}(E_\nu) = \Omega \, \varepsilon(E_\nu) N_\mathrm{nuc} / n}$, with $\Omega$ the solid angle to which the experiment is sensitive and $n$ the nucleon number density.
    
    \item $\Delta t$ is the total observation time.

    \item $E_\nu$ is the true neutrino energy.

    \item $\mathrm{d}\Omega = \mathrm{d}\varphi \, \mathrm{d}\cos \theta$, with $\varphi$ and $\theta$ the true azimuth and zenith angle of the neutrino arrival direction.

    \item $\frac{\mathrm{d}\phi}{\mathrm{d}E_\nu \mathrm{d}\Omega \mathrm{d}t \mathrm{d}A}$ is the UHE neutrino flux. We parameterize it as an isotropic power law,
    \begin{equation}
        \frac{\mathrm{d}\phi}{\mathrm{d}E_\nu \mathrm{d}\Omega \mathrm{d}t \mathrm{d}A} = \phi_0 \left(\frac{E_\nu}{E_\nu^\mathrm{ref}}\right)^{-\gamma}
    \end{equation}
    with $\phi_0$ the normalization, $E_\nu^\mathrm{ref}$ an arbitrary reference energy, and $\gamma$ the spectral index. Soft fluxes correspond to large values of $\gamma$, and hard fluxes to small $\gamma$ (soft fluxes fall more rapidly with energy than hard fluxes, as they have more low-energy or ``soft'' events).
    
    \item $\sigma(E_\nu)$ is the neutrino-nucleon interaction cross section.
    
    \item $nL(\theta)$ is the traversed chord weighted by the number of nucleons. Explicitly, it is
    \begin{equation}
        nL(\theta) = \int \mathrm{d}s \, n(s, \theta) \, ,
        \label{eq:grammage}
    \end{equation}
    with $\mathrm{d}s$ the length element along the chord and $n$ the nucleon number density.
    
    \item $R(E_\mathrm{rec}, E_\nu)$ is the energy reconstruction function, i.e., the probability to reconstruct an energy $E_\mathrm{rec}$ if the true energy is $E_\nu$. We assume it to be a Gaussian,
    \begin{equation}
        R(E_\mathrm{rec}, E_\nu) \propto \exp\left[-\frac{(\log_{10} E_\mathrm{rec} - \log_{10} E_\nu)^2}{2 (\Delta \log_{10} E_\nu)^2}\right] \, ,
    \end{equation}
    normalized such that $\int_0^\infty \mathrm{d}E_\mathrm{rec} \, R(E_\mathrm{rec}, E_\nu) = 1$. Here, $\Delta \log_{10} E_\nu$ is the energy resolution.

    \item $R(\theta_\mathrm{rec}, \theta)$ is the angle reconstruction function, i.e., the probability to reconstruct a zenith angle $\theta_\mathrm{rec}$ if the true zenith angle is $\theta$. We assume it to be a Gaussian,
    \begin{equation}
        R(\theta_\mathrm{rec}, \theta) \propto  \exp \left[-\frac{(\theta_\mathrm{rec} - \theta)^2}{2 (\Delta \theta)^2}\right] \, ,
    \end{equation}
    normalized such that $\int_0^{180^\circ} \mathrm{d}\theta_\mathrm{rec} \, R(\theta_\mathrm{rec}, \theta) = 1$. Here, $\Delta \theta$ is the angular resolution.

    \item $\varepsilon(E_\nu)$ is the detection efficiency, given by \cref{eq:logistic}. This includes trigger and analysis efficiencies (see Ref.~\cite{ARA:2022rwq}).

\end{itemize}
For computational simplicity, in our fits we use $\log_{10}E_\nu$ as a variable instead of $E_\nu$ (then $\mathrm{d}\phi/\mathrm{d}\log_{10}E_\nu \propto E_\nu^{-\gamma+1}$) and, instead of computing \cref{eq:expected} for all values of $E_\mathrm{rec}$ and $\theta_\mathrm{rec}$, we compute it in a fine 2-D grid and then interpolate. This also allows us to replace the integrations over $E_\nu$ and $\theta$ with array convolutions.

We then build an unbinned maximum likelihood,
\begin{equation}
    -2 \ln L (\gamma, \phi_0, S \equiv \sigma/ \sigma_\mathrm{SM})  = -2 \sum_i \ln \mu(E_\mathrm{rec}^i, \theta_\mathrm{rec}^i) \, 
\end{equation}
with $E_\mathrm{rec}^i$ and $\theta_\mathrm{rec}^i$ the reconstructed energy and angle of each event $i$, respectively. These are drawn from the 2-D probability distribution given by \cref{eq:expected} with $\sigma=\sigma_\mathrm{SM}$ and $\gamma=2.5$. The 1$\sigma$ allowed region on $S$ is then given by the values that satisfy
\begin{equation}
     \min_{\gamma, \phi_0} \left[-2 \ln L (\gamma, \phi_0, S) \right] - \min_{\gamma, \phi_0, S} \left[-2 \ln L (\gamma, \phi_0, S)\right] \leq 1 \, .
\end{equation}
Because $\phi_0$ is just a multiplicative constant, minimizing over it is trivial. We repeat this procedure, drawing $E_\mathrm{rec}^i$ and $\theta_\mathrm{rec}^i$ many times, to take into account the Poisson fluctuations of low-statistics data, and in our results we show the median 1$\sigma$ allowed region.

We have not included systematic uncertainties in our main results, as meeting the astrophysical goals of these detectors will require in any case systematics to be under control. Since systematic uncertainties affect different detectors in different ways, they are still under active investigation. For example, in-ice radio detectors have uncertainties in direction reconstruction owing to the complexities of radio propagation in polar ice, an area of active research~\cite{Besson:2021wmj, Allison:2019rgg, Jordan:2019bqu, Connolly:2021cum, Aguilar:2022kgi, RadarEchoTelescope:2020nhe}. In addition, recent studies~\cite{ARIANNA:2020zrg} observe systematic reconstruction offsets in RF arrival direction on the order of $0.1$ to $1$ degree. 

We have checked that an \emph{overall} systematic offset in $\theta$ does not affect our results, as it just moves the location of the horizon while keeping the \emph{slope} of the angle-dependent attenuation equal; the latter is what measures $\sigma$ (see \cref{fig:survival_1}). More generally, our results show that for neutrino energies $10^{7} \, \mathrm{GeV} \lesssim E_\nu \lesssim 10^9 \, \mathrm{GeV}$, $\sigma$ is mostly measured from the event asymmetry between $\theta \lesssim 91^\circ$ and $\theta \gtrsim 91^\circ$ (see \cref{fig:evt_distribution}); systematic angular offsets that do not affect the size of this asymmetry should not significantly affect sensitivity to $\sigma$.


\section{Sensitivity at different neutrino energies}
\label{appendix:energies}   

\begin{figure}[hbtp]
    \includegraphics[width=0.49\textwidth]{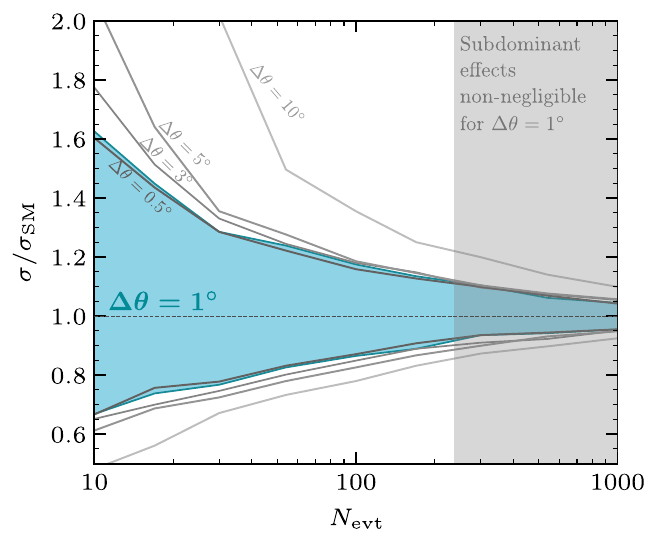} \hfill  \includegraphics[width=0.49\textwidth]{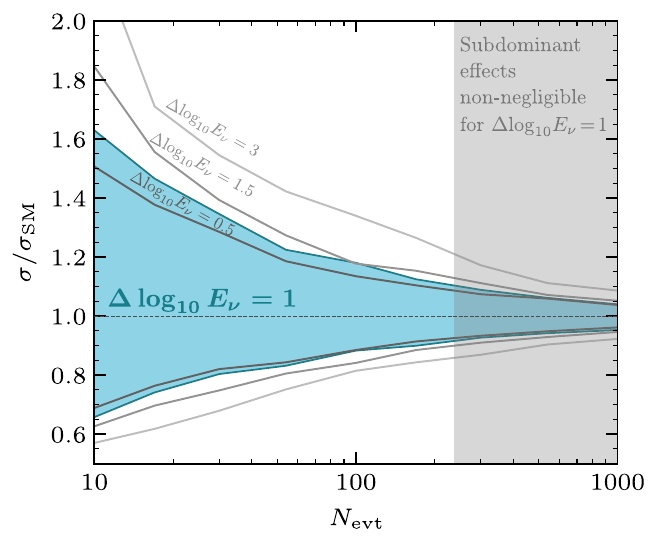}   
    \caption{Sensitivity to the cross section, as in \cref{fig:resolution} but for neutrino energies around $E_0 = 10^{7.5} \, \mathrm{GeV}$. In the left panel we take $\Delta \log_{10} E_\nu = 1$, and in the right panel $\Delta \theta = 1^\circ$. \emph{At lower energies, the angular resolution requirement is less stringent.}}
    \label{fig:resolution_75}
\end{figure}

\begin{figure}[hbtp]
    \includegraphics[width=0.49\textwidth]{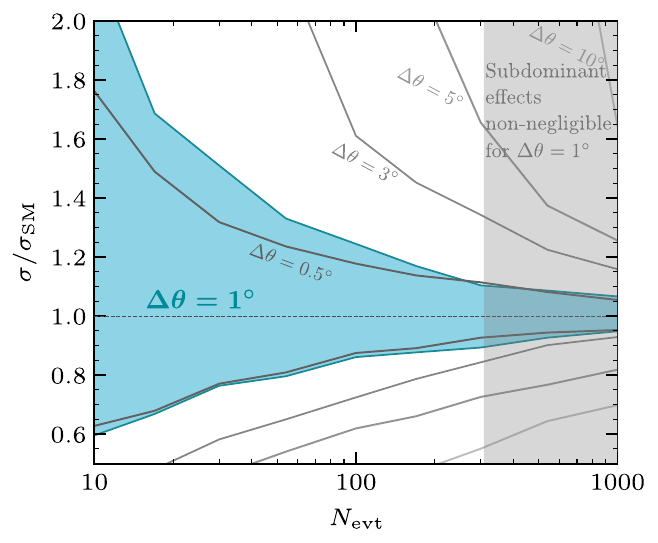} \hfill  \includegraphics[width=0.49\textwidth]{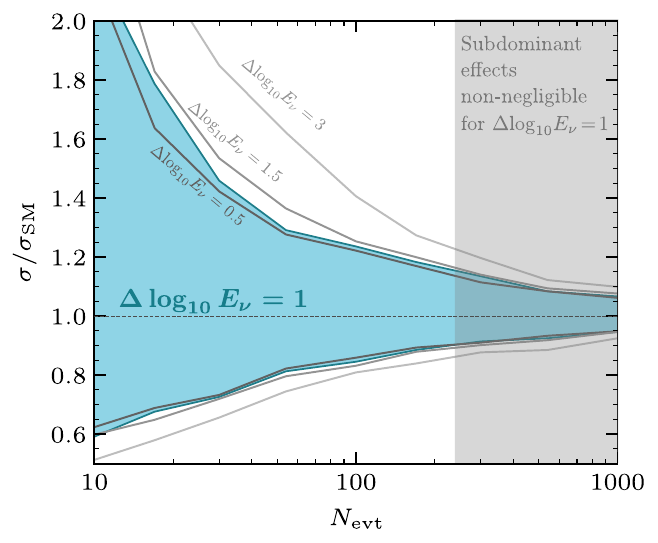}  
    \caption{Sensitivity to the cross section, as in \cref{fig:resolution} but for neutrino energies around $E_0 = 10^{9.5} \, \mathrm{GeV}$. In the left panel we take $\Delta \log_{10} E_\nu = 1$, and in the right panel $\Delta \theta = 1^\circ$. \emph{At higher energies, the angular resolution requirement is more stringent.}}
    \label{fig:resolution_95}
\end{figure}

In the main text, we assume for concreteness that the detector peak sensitivity is around $E_0 = 10^{8.5} \, \mathrm{GeV}$. \Cref{fig:resolution_75,fig:resolution_95} show that the benchmark resolutions at other energies are comparable. The sensitivities are not very different either, although, in general, for higher neutrino energies the measurement is more challenging due to the steeper angular distribution (see \cref{fig:survival_2}).

\begin{figure}[hbtp]
    \includegraphics[width=0.49\textwidth]{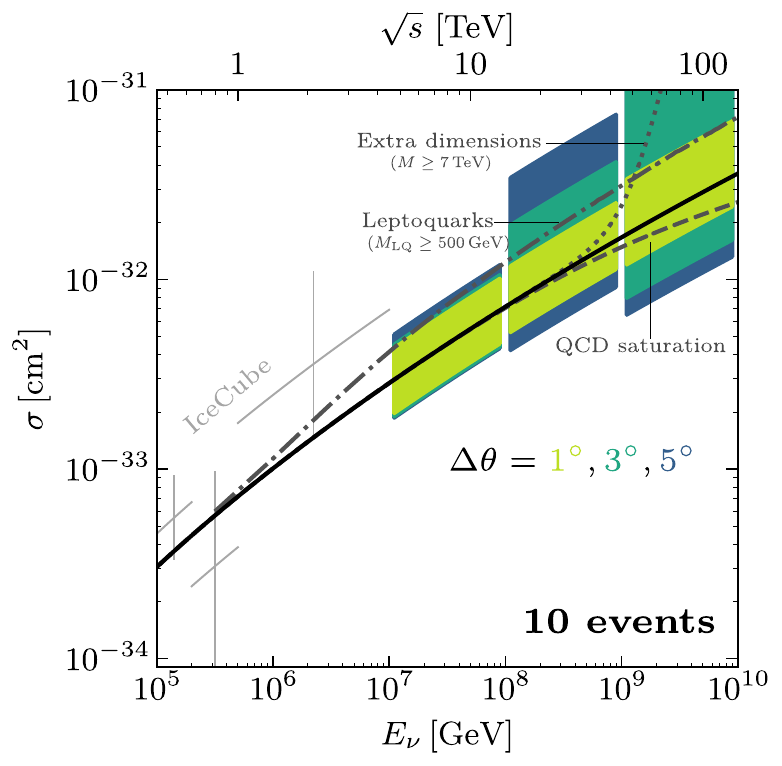} \hfill  \includegraphics[width=0.49\textwidth]{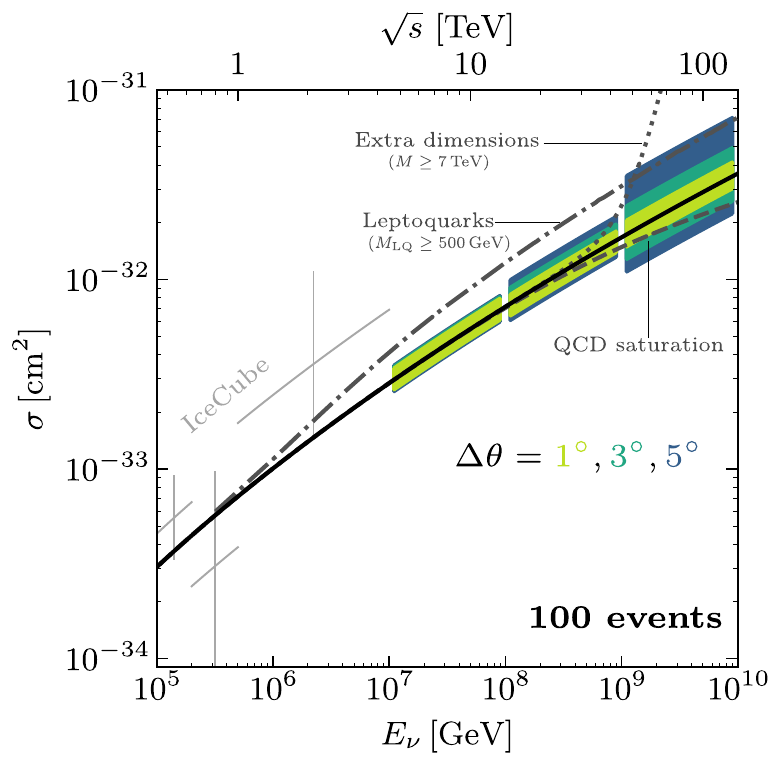}  
    \caption{Sensitivity to $\sigma$ as in \cref{fig:sensitivity_xsec} but for different angular resolutions. We take benchmark energy resolution $\Delta \log_{10} E_\nu = 1$.}
    \label{fig:sensitivity_xsec_dTheta}
\end{figure}

\Cref{fig:sensitivity_xsec_dTheta} further illustrates that angular resolution is more critical at higher energies, particularly for low statistics. Here, we assume our benchmark energy resolution, $\Delta \log_{10} E_\nu = 1$.

\section{Impact of the spectral slope}
\label{appendix:flux}

In the main text, we assume for concreteness that the true neutrino flux is given by $\mathrm{d}\phi / \mathrm{d}E_\nu \propto E_\nu^{-2.5}$ (although when fitting we marginalize over the spectral index). A power law is generically predicted by astrophysical models~\cite{Fermi:1949ee, Gaisser:2016uoy, Megnin:2000}, and is also phenomenologically justified due to the relatively small expected statistics and energy range. However, our choice of the true spectral index is \emph{a priori} arbitrary.

\begin{figure}[hbtp]
    \includegraphics[width=0.45\textwidth]{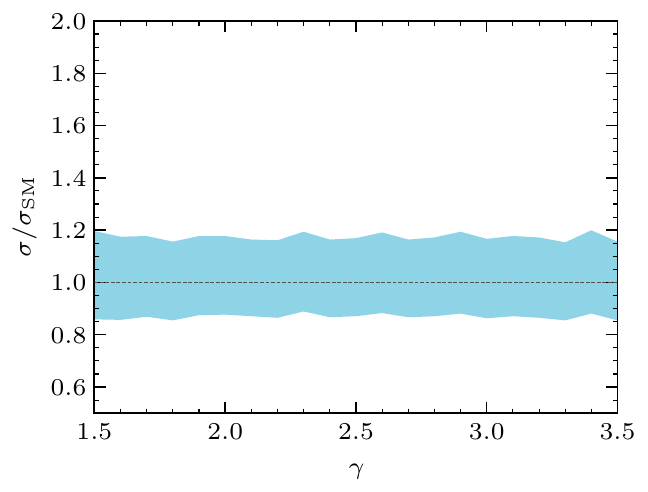}
    \caption{Sensitivity to $\sigma$ for different fluxes $\mathrm{d}\phi / \mathrm{d}E_\nu \propto E_\nu^{-\gamma}$. \emph{The sensitivity does not depend on the assumed spectral index, although it is important to marginalize over it when fitting.}}
    \label{fig:gamma}
\end{figure}

\Cref{fig:gamma} shows that the results are insensitive to the assumed true spectral index. We take our benchmark values $\Delta \theta = 1^\circ$ and $\Delta \log_{10} E_\nu = 1$, plus we set $N_\mathrm{evt} = 100$. The fluctuations are compatible with Monte-Carlo noise.

\section{Impact of the Earth profile}
\label{appendix:ice}

In the main text, we assume the PREM Earth density profile together with a 3-km ice layer. Here, we investigate how variations in these affect cross-section measurements.

\begin{figure}[hbtp]
    \centering
    \includegraphics[width=0.5\textwidth]{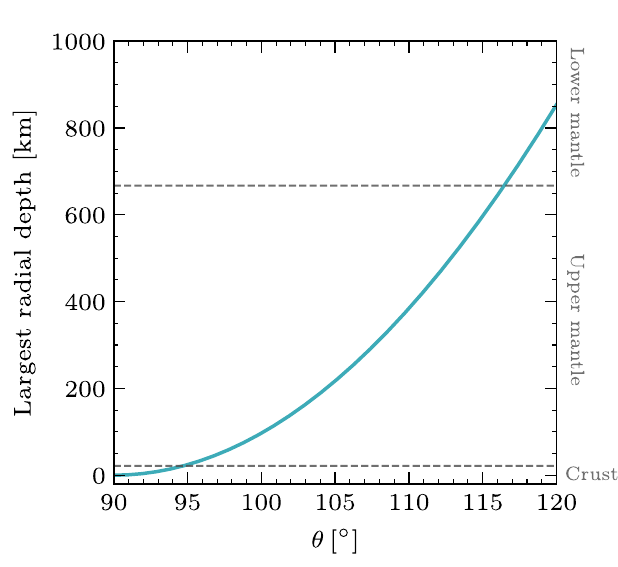}
    \caption{Largest depth of the neutrino trajectory at a given zenith angle $\theta$ for a detector on the Earth surface. \emph{UHE neutrino attenuation mostly probes the crust and upper mantle.}}
    \label{fig:depth}
\end{figure}

\Cref{fig:depth} shows that neutrino trajectories with zenith angles $\lesssim 110^\circ$ (these are the relevant angles at UHE energies, see \cref{fig:survival_2}) mostly cross the Earth crust and upper mantle. The Earth density there is well understood~\cite{Shen:2016kxw, Kai:2018, Megnin:2000, Laske:2013, Fretwell:2013, Weisen:2018, Bakhti:2020tcj}; moreover, neutrino attenuation is only sensitive to the \emph{integrated} density profile, i.e., to the \emph{average} density along the traversed chord (see \cref{eq:grammage}). 

\begin{figure}[hbtp]
    \centering
    \includegraphics[width=0.5\textwidth]{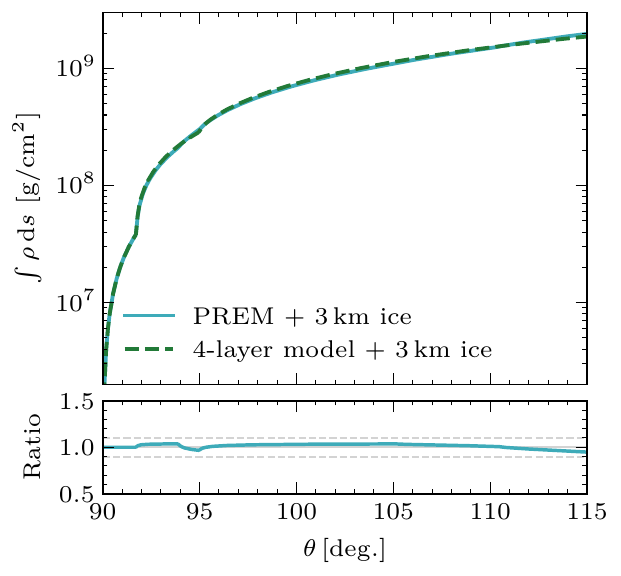}
    \caption{Grammage of neutrino trajectories with different zenith angles for different Earth models (see text). \emph{As neutrino attenuation is sensitive to the \emph{average} traversed density, our results are not sensitive to the details of the density profile}.}
    \label{fig:grammage}
\end{figure}

\Cref{fig:grammage} further illustrates that measuring the cross section does not require precise knowledge of the Earth density profile. The top panel shows the density-weighted traversed chord, or grammage, that controls Earth attenuation (see \cref{eq:grammage}) for different neutrino trajectories assuming the PREM profile [solid] or a simplified profile where the Earth is modeled by 4 layers of uniform density [dashed]: a 21.4-km thick crust with $\rho = 2.7 \, \mathrm{g}/\mathrm{cm}^3$, a 645.6-km thick upper mantle with $\rho = 3.5 \, \mathrm{g}/\mathrm{cm}^3$, a 2221-km thick lower mantle with $\rho = 5 \, \mathrm{g}/\mathrm{cm}^3$, and a core with a radius of 3480 km and $\rho = 11 \, \mathrm{g}/\mathrm{cm}^3$. In all cases, we include a 3-km ice layer. The error introduced by modeling the Earth as a few uniform-density layers is smaller than the $\sim 10\%$ subdominant effects in $\sigma$ that we ignore throughout this paper.

\begin{figure}[hbtp]
    \includegraphics[width=0.45\textwidth]{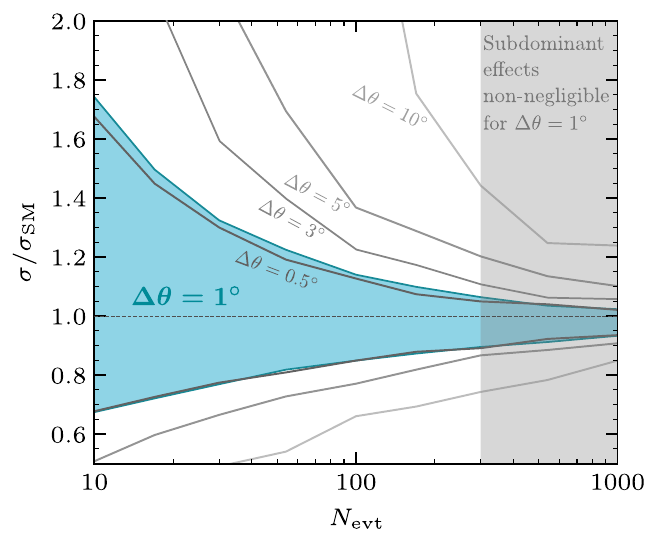}
    \caption{Sensitivity to the cross section, as in \cref{fig:resolution} but without including an ice layer. We take benchmark $\Delta \log_{10} E_\nu = 1$ and neutrino energies around $E_0 = 10^{8.5} \, \mathrm{GeV}$. \emph{The sensitivity is independent of the specific density profile below the detector.}}
    \label{fig:resolution_ice}
\end{figure}

Finally, \cref{fig:resolution_ice} shows that if we do not include the 3-km ice layer the sensitivity to $\sigma$ is not appreciably modified. This indicates that our conclusions are robust with respect to the Earth density profile at the location of the detector, as long as the analysis is carried out assuming the correct profile. We have also checked that, if the ice layer is present, the results do not appreciably depend on the detector being buried within the ice, as expected because the neutrino absorption length is much larger than the ice layer depth. 

\clearpage
\newpage
\section{Neutrino regeneration in Earth}
\label{appendix:tau}

In the main text, we do not include neutrino regeneration in Earth. In principle, even though UHE neutrinos are attenuated by the Earth, neutral-current and $\nu_\tau$ charged-current interactions produce secondary neutrino fluxes at lower energies~\cite{Ritz:1987mh, Nicolaidis:1996qu, Halzen:1998be, Kwiecinski:1998yf, Beacom:2001xn, Dutta:2002zc, Soto:2021vdc, Arguelles:2022aum}. This could partially compensate attenuation. However, due to the steeply falling flux as a function of neutrino energy, the regenerated flux is generically subdominant to the primary flux at lower energies. 

If this effect were to be important, $\sigma$ could not be measured in a model-independent way. Regeneration would be affected by the identity and kinematics of the produced particles in any new interaction, and the analysis would have to be performed on a model-by-model basis.

To quantify the importance of regeneration, we have evolved a $\nu_\tau$ flux (the flavor for which regeneration is maximal) using the public software \verb+TauRunner+~\cite{Safa:2019ege, Safa:2021ghs}.  

\begin{figure}[hbtp]
    \includegraphics[width=0.329\textwidth]{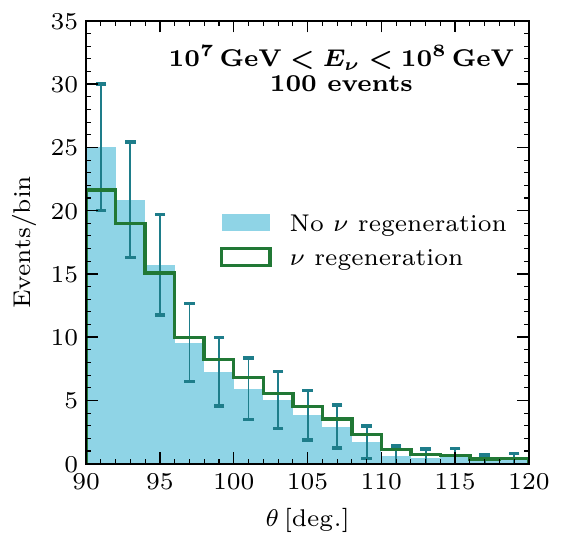} \includegraphics[width=0.329\textwidth]{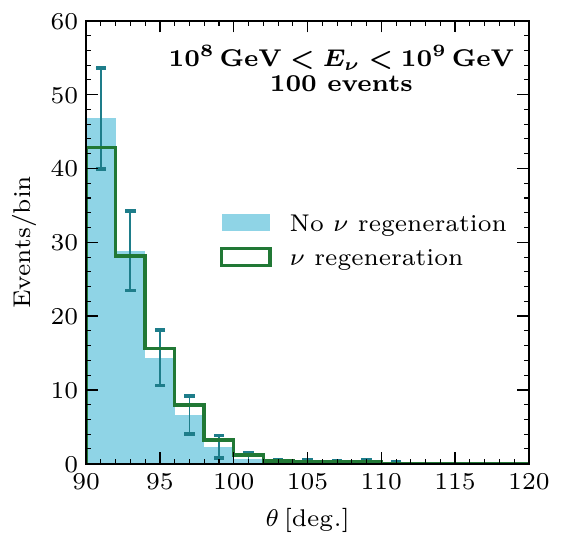} \includegraphics[width=0.329\textwidth]{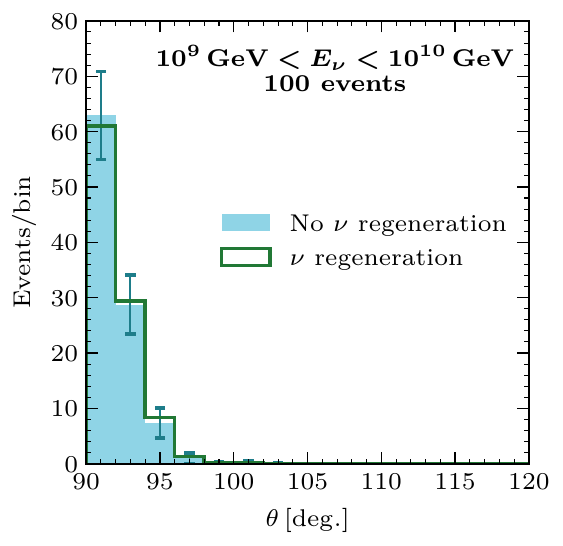}   
    \caption{Normalized angular distributions at different energies for a $\nu_\tau$ flux $\mathrm{d}\phi / \mathrm{d}E_\nu \propto E_\nu^{-2.5}$ propagating through the Earth. Error bars show the expected statistical uncertainties if 100 events are observed. For clarity, we do not include energy-dependent detection efficiency effects. \emph{Unless high statistics are reached, regeneration effects are subdominant.}}
    \label{fig:regeneration}
\end{figure}

\Cref{fig:regeneration} shows that regeneration effects are subdominant. The induced deviations  in the normalized angular distributions are $\lesssim 10 \%$, well below statistical uncertainties unless the number of events is very large. Regeneration effects may be important if, for a large energy range, the spectrum is not steeply falling (e.g., $\mathrm{d}\phi / \mathrm{d}E_\nu \propto E_\nu^{-1}$). Some cosmogenic neutrino models predict such spectra~\cite{Romero-Wolf:2017xqe, AlvesBatista:2018zui, Heinze:2019jou} but only for small energy ranges.

\end{document}